\documentclass[12pt]{article}
\usepackage[a4paper]{geometry}
\usepackage[english]{babel}

\usepackage{amsmath}
\usepackage{amsfonts}
\usepackage{amstext}
\usepackage{amssymb}
\usepackage{amsthm}
\usepackage{amscd}

\usepackage{xcolor}
\definecolor{myurlcolor}{rgb}{0,0.5,0}
\definecolor{mycitecolor}{rgb}{0,0,1}
\definecolor{myrefcolor}{rgb}{0.5,0,0}
\usepackage{graphicx}
\usepackage{tikz}
\usepackage{tikz-cd}

\usepackage[pagebackref,draft=false]{hyperref}
\hypersetup{colorlinks,
linkcolor=myrefcolor,
citecolor=mycitecolor,
urlcolor=myurlcolor}

\usepackage[capitalize]{cleveref}
\usepackage{caption}

\usepackage{etoolbox}
\usepackage{makeidx}
\usepackage{sectsty}
\usepackage{mathrsfs}
\usepackage{dsfont}
\usepackage{enumitem} 
\usepackage[]{latexsym}
\usepackage{braket}
\usepackage{caption}

\newtheorem{remark}{Remark}

\newtheorem{theorem}{Theorem}

\newtheorem{proposition}{Proposition}

\newtheorem*{proof*}{Proof}
\newtheorem*{proposition*}{Proposition}

\newcommand{\be}{\begin{equation}}
\newcommand{\ee}{\end{equation}}
\newcommand{\bea}{\begin{eqnarray}}
\newcommand{\eea}{\end{eqnarray}}
\newcommand{\vsp}{\vspace{0.4cm}}
\newcommand{\grit}[1]{{\bfseries {\itshape {#1}}}}

\newcommand{\ra}{\rightarrow}

\newcommand{\lra}{\longrightarrow}

\newcommand{\hh}{\mathcal{H}}
\newcommand{\bh}{\mathcal{B}(\mathcal{H})}

\newcommand{\Uh}{\mathcal{U}(\mathcal{H})}

\newcommand{\Tr}{\textit{Tr}}

\newcommand{\stsp}{\mathscr{S}}
\newcommand{\stspn}{\mathscr{N}}

\newcommand{\appa}{\mathscr{A}}

\newcommand{\gapp}{\mathscr{G}}
\newcommand{\uapp}{\mathscr{U}}

\newcommand{\posn}{\mathscr{P}_{*}}
\newcommand{\pposn}{(\mathscr{P}_{*})}

\newcommand{\gr}{\mathrm{g}}

\title{Manifolds of classical probability distributions and quantum density operators in infinite dimensions}

\author{F. M. Ciaglia$^{1,6}$\href{https://orcid.org/0000-0002-8987-1181}{\includegraphics[scale=0.7]{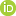}}, A. Ibort$^{2,3,7}$\href{https://orcid.org/0000-0002-0580-5858}{\includegraphics[scale=0.7]{ORCID.png}}, J. Jost$^{1,8}$\href{https://orcid.org/0000-0001-5258-6590}{\includegraphics[scale=0.7]{ORCID.png}}, G. Marmo$^{4,5,9}$\href{https://orcid.org/0000-0003-2662-2193}{\includegraphics[scale=0.7]{ORCID.png}} \\
\footnotesize{$^{1}$\textit{ Max Planck Institute for Mathematics in the Sciences, Leipzig, Germany}} \\
\footnotesize{$^{2}$\textit{ ICMAT, Instituto de Ciencias Matem\'{a}ticas (CSIC-UAM-UC3M-UCM)}}  \\
\footnotesize{$^{3}$\textit{ Depto. de Matem\'aticas, Univ. Carlos III de Madrid, Legan\'es, Madrid, Spain}}  \\
\footnotesize{$^{4}$\textit{ Dipartimento di Fisica ``E. Pancini'', Universit\`a di Napoli Federico II, Napoli, Italy}} \\
\footnotesize{$^{5}$\textit{ INFN-Sezione di Napoli, Napoli, Italy.}} \\
\footnotesize{$^{6}$\textit{ e-mail: \texttt{florio.m.ciaglia[at]gmail.com}}, $^{7}$\textit{ e-mail: \texttt{albertoi[at]math.uc3m.es}}} \\
\footnotesize{$^{8}$\textit{ e-mail: \texttt{jjost[at]mis.mpg.de}}, $^{9}$\textit{ e-mail: \texttt{marmo[at]na.infn.it}}}
}

\date{}

\begin{document}

\maketitle

\begin{abstract}
The manifold structure of subsets of classical probability distributions and quantum density operators in infinite dimensions is investigated in the context of $C^{*}$-algebras and actions of  Banach-Lie groups.
Specificaly, classical probability distributions and quantum density operators may be both described as states (in the functional analytic sense) on a given $C^{*}$-algebra $\appa$ which is Abelian for Classical states, and non-Abelian for Quantum states.
In this contribution, the space  of states $\stsp$ of a possibly infinite-dimensional, unital $C^{*}$-algebra $\appa$ is partitioned into the disjoint union of the orbits of an action of the group $\gapp$ of invertible elements of $\appa$.
Then, we prove that the orbits through density operators on an infinite-dimensional, separable Hilbert space $\hh$  are smooth, homogeneous Banach manifolds of $\gapp=\mathcal{GL}(\hh)$, and, when $\appa$ admits a faithful tracial state $\tau$ like it happens in the Classical case when we consider probability distributions with full support, we prove that the orbit through $\tau$ is a smooth, homogeneous Banach manifold for $\gapp$.

\end{abstract}

\tableofcontents

\section{Introduction}

The use of differential geometric methods in the context of classical and quantum information theory is a well-established and flourishing trend. 
This has led to the birth of new perspectives in the understanding of theoretical issues, as well as to numerous achievements in the realm of applications.
At the heart of this methodological attitude towards classical and quantum information geometry there is the  notion of a smooth manifold.
This clearly follows from the fact that differential geometry deals with smooth manifolds and with all the additional structures with which smooth manifolds may be dressed.
However, the smooth manifolds employed in the vast majority of the literature pertaining to classical and quantum information geometry are finite-dimensional.
This is essentially due to the fact that working with infinite-dimensional manifolds requires  to carefully handle a nontrivial number of technical issues, and these technicalities may obscure the conceptual ideas one wants to convey.
Consequently, it has been, and it still is useful to focus on finite-dimensional systems in order to explicitly develop new ideas, and to postpone the analysis of the infinite-dimensional systems to later times.
On the other hand, the number of conceptual results on finite-dimensional systems is growing so rapidly that we may dare to say to have a well-estabilished theoretical backbone for the information geometry of finite-dimensional systems so that it is reasonable to start looking in more detail at the infinite-dimensional systems.

\vsp

Of course, there already have been contributions in the information geometry of infinite-dimensional systems.
For instance, in \cite{pistone_sempi-an_infinite_dimensional_geometric_structure_on_the_space_of_all_the_probability_measures_equivalent_to_a_given_one}, a Banach manifold structure is given to the set $M_{\mu}$ of all probability measures on some measure space $(\mathcal{X},\Sigma)$ that are mutually absolutely continuous with respect to a given probability measure $\mu$ on $(\mathcal{X},\Sigma)$ by means of Orlicz spaces, and, in \cite{gibilisco_pistone-connections_on_nonparametric_statistical_manifolds_by_orlicz_space_geometry}, the infinite-dimensional analogue of the $\alpha$-connections of Amari and Cencov on this class of manifolds is studied.
Orlicz spaces were also employed in the quantum framework  in \cite{grasselli_streater-the_quantum_information_manifold_for_epsilon-bounded_froms,streater-quantum_orlicz_spaces_in_information_geometry} to build a Banach manifold structure on Gibbs-like density operators on an infinite-dimensional, complex, separable Hilbert space $\hh$, and in \cite{jencova-affine_connections_duality_and_divergences_for_a_von_neumann_algebra,jencova-a_construction_of_a_nonparametric_quantum_information_manifold} to build a Banach manifold structure on the space of faithful, normal states on an abstract von Neumann algebra.

In \cite{friedrich}, a Hilbert manifold was obtained by equipping the space of $L^2$-probability measures with the Fisher metric.
In \cite{newton-an_infinite_dimensional_statistical_manifold_modelled_on_hilbert_space}, a Hilbert manifold structure is given to a subset of $M_{\mu}$ characterized by some constraint relations, and the $\alpha$-connections on it are studied.
In \cite{bauer_bruveris_michor-uniqueness_of_the_fisher-rao_metric_on_the_space_of_smooth_densities}, the uniqueness (up to rescaling) of the Fisher-Rao metric tensor on the Frechet manifold $\mathrm{Prob}(M)$ of smooth positive densities normalized to $1$ on a smooth, compact manifold $M$ under the requirement of invariance with respect to the group of diffeomorphisms of $M$ is solved.

In \cite{ay_jost_vanle_schwachhofer-information_geometry_and_sufficient_statistics,ay_jost_vanle_schwachhofer-information_geometry,ay_jost_vanle_schwachhofer-parametrized_measure_models}, a new approach to infinite-dimensional parametric models of probability measures on some measure space is taken, and tensorial structures are obtained by exploiting the natural immersion of the space of probability measures into the Banach space of signed finite measures (where the norm is given by the total variation).
This makes the theory independent of the choice of a reference measure, as everything transforms appropriately and integrability conditions are preserved when the reference measure is changed.

From a different point of view, the structure of infinite dimensional groups has been treated exhaustively in relation with mathematical physics problems, like hydrodynamical-like equations for instance, involving probability densities.  It was realized that the proper way to deal with such problems was to consider a weaker form of differentiability called IHL-Lie groups introduced by Omori \cite{omori} (see for instance \cite{adams_ratiu} and references therein).

In the context of quantum information theory, the geometrization of some of the relevant structures, for instance the K\"{a}hler-Hilbert manifold structure on the space of pure quantum states given by  the complex projective space of an infinite-dimensional, complex, separable Hilbert space $\mathcal{H}$, together with a Hamiltonian formulation of the unitary evolutions of quantum mechanics, as given for instance in \cite{ashtekar_schilling-geometrical_formulation_of_quantum_mechanics,cirelli_lanzavecchia_mania-normal_pure_states_and_the_von_neumann_algebra_of_bounded_operators_as_kahler_manifold,cirelli_mania_pizzocchero-quantum_mechanics_as_an_infinite_dimensional_Hamiltonian_system_with_uncertainty_structure,cirelli_pizzocchero-on_the_integrability_of_quantum_mechanics_as_an_infinite_dimensional_system,kibble-geometrization_of_quantum_mechanics}, allows a simpler treatment of the differentiable structures of the corresponding infinite-dimensional groups present in the theory  
as it is shown in \cite{andruchow_stojanoff-differentiable_structure_of_similarity_orbits,andruchow_stojanoff-geometry_of_unitary_orbits,andruchow_varela-Riemannian_geometry_of_finite_rank_positive_operators,beltita_ratiu-symplectic_leaves_in_real_banach_lie-poisson_spaces,bona-some_considerations_on_topologies_of_infinite_dimensional_unitary_coadjoint_orbits,grabowski_kus_marmo_shulman-geometry_of_quantum_dynamics_in_infinite_dimensional_hilbert_space,larotonda-the_metric_geometry_of_infinite_dimensional_Lie_groups_and_their_homogeneous_spaces}, where the action of Banach-Lie groups of unitary operators on an infinite-dimensional, complex, separable Hilbert space $\mathcal{H}$ is used to give a Banach manifold structure to appropriate subsets of quantum density operators, positive semidefinite linear operators and elements of Banach Lie-Poisson spaces or, as it will be shown in this paper, to certain orbits of the group of invertible elements on a $C^*$-algebra.

\vsp

The purpose of this contribution is to look at infinite-dimensional systems in both classical and quantum information geometry from the unifying perspective coming from  the interplay between the theory of $C^*$-algebras and the infinite-dimensional differential geometry of Banach manifolds and Banach-Lie groups.
The choice of $C^{*}$-algebras as a main ingredient is due to the fact that classical spaces of probability distributions as well as spaces of quantum states are both concrete realizations of the same mathematical object, i.e., the space $\stsp$ of (mathematical) states on a $C^{*}$-algebra $\appa$, with the classical case characterized by the requirement that $\appa$ is Abelian.

Let us explain the motivations behind our idea by looking at a finite-dimensional example.
Consider a quantum system described by a complex Hilbert space $\hh$ with $\mathrm{dim}(\hh)=N<\infty$.
According to the formalism of standard quantum mechanics, a (bounded) observable $\mathbf{a}$ of the system  is an element of the algebra $\bh$ of bounded, linear operators on $\hh$, while a state $\rho$ of the system is a positive linear functional on $\hh$ such that $\rho(\mathbb{I})=1$, where $\mathbb{I}\in\bh$ is the identity operator.
Since $\mathrm{dim}(\hh)=N<\infty$, we may identify the dual space of $\bh$ with $\bh$ itself by means of the trace operation $\Tr$ on $\hh$, that is, an element $\xi\in\bh$ determines a linear functional on $\bh$ by means of
\be
\xi(\mathbf{a})\,:=\,\Tr(\xi\,\mathbf{a}),
\ee
and every linear functional on $\bh$ is of this form.
Consequently, a quantum state may be identified with a so-called density operator on $\hh$, that is, a self-adjoint, positive semidefinite, linear operator $\rho$ such that $\Tr(\rho)=1$.
The set of all density operators on $\hh$ is denoted by $\stsp(\hh)$, and it is a convex body in the affine hyperplane $\mathfrak{T}_{1}(\hh)$ of self-adjoint, linear operators with unit trace.
The interior of $\stsp(\hh)$ in $\mathfrak{T}_{1}(\hh)$ is an open convex set made of invertible (full-rank) density operators and denoted by $\stsp_{N}(\hh)$.
Being an open set in the affine hyperplane $\mathfrak{T}_{1}(\hh)$, the set $\stsp_{N}(\hh)$ admits a natural structure of smooth manifold modelled on $\mathfrak{T}_{1}(\hh)$, and this manifold structure makes $\stsp_{N}(\hh)$ the subject of application of the methods of classical information geometry in the context of quantum information (see \cite{ay_tuschmann-duality_versus_flatness_in_quantum_information_geometry,gibilisco_isola-wigner-yanase_information_on_quantum_state_space:the_geometric_approach,hasegawa-noncommutative_extension_of_information_geometry,hasegawa_petz-noncommutative_extension_of_information_geometry_II,naudts-quantum_statistical_manifolds,petz_sudar-geometries_of_quantum_states}).
Note that, if we consider the subset $\stsp_{\rho}(\hh)$ of  density operators commuting with a fixed $\rho\in\stsp$ and mutually-commuting with each other, the spectral theorem assures us that $\stsp_{\rho}(\hh)$ may be identified with the $N$-dimensional simplex representing classical probability distributions on a finite sample space.

The manifold structure on $\stsp_{N}(\hh)$ is compatible with an action of the Lie group $\gapp(\hh)$ of invertible linear operators on $\stsp(\hh)$.
Specifically, if $\gr\in\gapp(\hh)$ and $\rho\in\stsp_{N}(\hh)$, we may define the map\footnote{More generally, this action is well-defined on the whole $\stsp(\hh)$ and its orbits are given by density operators with fixed rank (see \cite{chruscinski_ciaglia_ibort_marmo_ventriglia-stratified_manifold_of_quantum_states} for a recent review).} 
\be\label{eqn: nonlinear action density matrices}
(\gr,\,\rho)\,\mapsto\,\frac{\gr\,\rho\,\gr^{\dagger}}{\Tr(\gr\,\rho\,\gr^{\dagger})},
\ee
an this map defines a smooth, transitive action of $\gapp(\hh)$ on $\stsp_{N}(\hh)$.
In particular, if we consider the subgroup $\Uh$ of unitary operators,  we obtain the co-adjoint action $\mathbf{U}\,\rho\,\mathbf{U}^{\dagger}$ of $\Uh$ the orbits of which are density operators with fixed eigenvalues.
From this, it is clear that the manifold $\stsp_{N}(\hh)$ carries also the structure of homogeneous space of the Lie group $\gapp(\hh)$, and it is precisely this feature that we aim to extend to the infinite-dimensional setting.

\vsp

The paper is structured as follows.
In section \ref{sec: action}, given a possibly infinite-dimensional, unital $C^*$-algebra $\appa$, we will first define a linear action $\alpha$ of the Banach-Lie group  $\gapp$ of invertible elements of $\appa$ on the space $\appa^{*}_{sa}$ of self-adjoint linear functionals on $\appa$ that preserves the cone of positive linear functionals.
In subsection \ref{sebsec: positive trace-class operators}, we will analyse the case where $\appa$ is the algebra $\bh$ of bounded linear operators on a complex, separable Hilbert space $\hh$.
Specifically, we will prove that, if $\varrho$ is any positive trace-class operator on $\hh$ to which it is associated a unique normal, positive linear functional on $\bh$, the orbit of $\gapp=\mathcal{GL}(\hh)$ (bounded, invertible linear operators on $\hh$) through $\varrho$ by means of the linear action $\alpha$ is a homogeneous Banach manifold of $\gapp$.
Furthermore,  we provide sufficient conditions for two normal, positive linear functionals to belong to the same orbit of $\gapp$.
In subsection \ref{subsec: faithful, finite trace} we will prove that, if $\appa$ admits a faithful, finite trace $\tau$, the orbit through $\tau$ is a smooth, homogeneous Banach manifold of $\gapp$.

The action $\alpha$ does not preserve the space of states $\stsp$, and this leads us to present, in section \ref{sec: state-preserving action}, a ``deformation'' of $\alpha$, denoted by $\Phi$, which is an infinite-dimensional counterpart of the map given in equation \eqref{eqn: nonlinear action density matrices} and which is  a left action of $\gapp$ on the space $\stsp$ of states of $\appa$.
We prove that an orbit of $\gapp$ through $\rho\in\stsp$ by means of $\Phi$ is a homogeneous Banach manifold of $\gapp$ if and only if the orbit of $\gapp$ through $\rho$ by means of $\alpha$ is so.
We exploit this fact in subsection \ref{sec: density operators} where we apply the theory to the case $\appa=\bh$ with $\hh$ a complex, separable Hilbert space.
In particular, we obtain that the space of normal states on $\bh$, which can be identified with the space of density operators on $\hh$, is partitioned into the disjoint union of homogeneous Banach manifolds of $\gapp$, and, as we do for the case of normal, positive linear functionals, we provide sufficient conditions for two normal states to belong to the same orbit of $\gapp$.
In this context, when $\hh$ is infinite-dimensional, the space of faithful, normal states on $\bh$ may not be identified with a  convex, open submanifold of the space of self-adjoint, linear operators on $\hh$  with unit trace as it happens in the finite-dimensional case.
The results we present point out that, if we consider the manifold structure to be associated with a not-necessarily-convex group action, any faithful, normal state on $\bh$ lies on a smooth homogeneous Banach manifold which is an orbit of $\gapp$ by means of $\Phi$.
However, it is still an open question if there is only one such orbit for faithful, normal states.
In subsection \ref{subsec: faitfhul, tracial state} we consider  the case where $\appa$ admits a faithful, tracial state $\tau$, and we obtain that the orbit through $\tau$ is a smooth, homogeneous Banach manifold for $\gapp$.

Some concluding remarks are presented in section \ref{sec: concluding remarks}, while appendix \ref{app: C*-algebras preliminaries} is devoted to a brief introduction of the main notions, results and definitions concerning the theory of $C^{*}$-algebras for which a more detailed account can be found in \cite{blackadar-operator_algebras_theory_of_c*-algebras_and_von_neumann_algebras,bratteli_robinson-operator_algebras_and_quantum_statistical_mechanics_1,kadison_ringrose-fundamentals_of_the_theory_of_operator_algebras_I,sakai-C_star_and_W_star_algebras,takesaki-theory_of_operator_algebra_I}.
In appendix \ref{app: banach-lie groups and homogeneous spaces} we recall some notions, results and definitions concerning Banach-Lie groups and their homogeneous spaces.
In this case, we refer to \cite{abraham_marsden_ratiu-manifolds_tensor_analysis_and_applications,bourbaki-groupes_et_algebres_de_lie,chu-jordan_structures_in_geometry_and_analysis,lang-fundamentals_of_differential_geometry,upmeier-symmetric_banach_manifolds_and_jordan_calgebras} for a detailed account of the infinite-dimensional formulation of differential geometry that is used in this paper.

\section{Positivity-preserving action of $\gapp$}\label{sec: action}

Let $\appa$ be a possibly infinite-dimensional, unital $C^{*}$-algebra, that is, a $C^{*}$-algebra with a multiplicative identity element denoted by $\mathbb{I}$.
The existence of an identity element $\mathbb{I}$ in $\appa$ allows us to define the set $\gapp$ of invertible elements in $\appa$, that is, the set of all $\gr\in\appa$ admitting an inverse $\gr^{-1}\in\appa$ such that $\gr\,\gr^{-1}\,=\,\mathbb{I}$.
This is an open subset of $\appa$, and, when endowed with the multiplication operation of $\appa$, it becomes  a real Banach-Lie group in the relative topology induced by the norm topology of $\appa$.   
The Lie algebra $\mathfrak{g}$ of $\gapp$ can be identified with $\appa$ which is itself a real Banach-Lie algebra  (see \cite[p. 96]{upmeier-symmetric_banach_manifolds_and_jordan_calgebras}).   
We may define also the subgroup of unitary elements $\mathbf{u} \in \gapp$ as those invertible elements such that $\mathbf{u}^* = \mathbf{u}^{-1}$.   
Then, denoting such subgroup by $\uapp$, we get that $\uapp \subset \gapp$ is a closed Banach-Lie subgroup.

The purpose of this paper is to show that some of the homogeneous spaces of the group $\gapp$ are actually subsets of the space of states $\stsp$ on $\appa$.
Accordingly, even if $\stsp$ lacks of a differential structure as a whole, we may partition it into the disjoint union of Banach manifolds that are homogeneous spaces of the Banach-Lie group $\gapp$.
In order to do this, we will first consider an action $\alpha$ of $\gapp$ on the space $\appa^{*}_{sa}$ of self-adjoint linear functionals on $\appa$.
This action is linear,  and preserves the positivity of self-adjoint linear functionals, and we show that the orbits inside the cone $\appa^{*}_{+}$ of positive linear functionals are homogeneous Banach manifolds of $\gapp$.
However, the action $\alpha$ does not preserve the space of states $\stsp$, and we need to suitably deform it in order to overcome this difficulty.
The resulting action, denoted by $\Phi$, is well-defined only on the space of states $\stsp$, and we will prove that the orbits of $\Phi$ are homogeneous Banach manifolds of $\gapp$.

We introduce a map $\widetilde{\alpha}\colon\appa\,\times\,\appa^{*}_{sa}\,\lra\,\appa^{*}_{sa}$ given by
\be\label{eqn: widetildealpha map}
(\mathbf{a},\,\xi)\,\mapsto\,\widetilde{\alpha}(\mathbf{a},\,\xi)\,:=\,\xi_{\mathbf{a}},\;\;\;\xi_{\mathbf{a}}(\mathbf{b})\,:=\,\xi(\mathbf{a}^{\dagger}\,\mathbf{ba})\;\,, \qquad \forall\mathbf{b}\in \appa_{sa} .
\ee
Clearly, this map is linear in $\xi$ and it is possible to prove that $\widetilde{\alpha}$ is smooth with respect to the \grit{real} Banach manifold structures of $\appa\,\times\,\appa^{*}_{sa}$ (endowed with the smooth structure which is the product of the smooth structures of $\appa$ and $\appa^{*}_{sa}$) and $\appa^{*}_{sa}$.

\begin{proposition}\label{prop: the map alpha is smooth}
The map $\widetilde{\alpha}\colon\appa\,\times\,\appa^{*}_{sa}\,\lra\,\appa^{*}_{sa}$ is smooth.

\proof
Given $\mathbf{a},\mathbf{b}\in\appa$ and  $\xi\in\appa^{*}_{sa}$ we define $\xi_{\mathbf{a}\mathbf{b}}\in\appa^{*}_{sa}$ to be 
\be
\xi_{\mathbf{a}\mathbf{b}}(\mathbf{c})\,:=\,\frac{1}{2}\,\left(\xi(\mathbf{a}^{\dagger}\,\mathbf{c}\,\mathbf{b}) + \xi(\mathbf{b}^{\dagger}\,\mathbf{c}\,\mathbf{a})\right)\;\;\forall\,\mathbf{c}\,\in\,\appa_{sa} .
\ee
Then, we consider the map $F\colon\,(\appa\times \appa^{*}_{sa})\times(\appa\times \appa^{*}_{sa})\times(\appa\times \appa^{*}_{sa})\rightarrow \appa^{*}_{sa}$ given by
\be
 F(\mathbf{a},\xi\,;\mathbf{b},\zeta\,;\mathbf{c},\vartheta)\,:=\,\frac{1}{3}\left(\xi_{\mathbf{bc}} + \zeta_{\mathbf{ca}} + \vartheta_{\mathbf{ab}}\right)\,.
\ee
A direct computation shows that $F$ is a bounded multilinear map and that
\be
\widetilde{\alpha}(\mathbf{a},\,\xi)\,=\,F(\mathbf{a},\xi\,;\mathbf{a},\xi\,;\mathbf{a},\xi)\,,
\ee
which means that $\widetilde{\alpha}$ is a continuous polynomial map between $\appa\times\appa^{*}_{sa}$ and $\appa^{*}_{sa}$, hence, it is smooth with respect to the \grit{real} Banach manifold structures of $\appa\times\appa^{*}_{sa}$ and $\appa^{*}_{sa}$ (see \cite[p. 63]{chu-jordan_structures_in_geometry_and_analysis}).
\qed

\end{proposition}

Since $\gapp$ is an open submanifold of $\appa$, the canonical immersion $i_{\gapp}\colon\gapp\lra\appa$ given by $i_{\gapp}(\gr)=\gr$ is smooth.
Consequently, we may define the map 
\be\label{eqn: linear action of invertible elements on the dual}
\begin{split}
\alpha\,&\,\colon\gapp\,\times\,\appa^{*}_{sa}\,\lra\,\appa^{*}_{sa} \\
\alpha\,&:=\,\widetilde{\alpha}\,\circ\,(i_{\gapp}\,\times\,\mathrm{id}_{\appa^{*}}),
\end{split}
\ee
where $\mathrm{id}_{\appa^{*}_{sa}}$ is the identity map, and this map is clearly smooth because it is the composition of smooth maps.

A direct computation shows that $\alpha$ is a (smooth) left action of $\gapp$ on $\appa^{*}_{sa}$.
We are interested in the orbits of $\alpha$, in particular, we are interested in the orbits passing through positive linear functionals.   It is possible to prove the following proposition.

\begin{proposition}\label{prop: useful properties of convex action}
Let $\alpha$ be the action of $\gapp$ on $\appa^{*}$, then we have

\begin{enumerate}
\item if $\appa$ is a $W^*$-algebra, then $\alpha$ preserves the space $(\appa_{*})_{sa}$ of self-adjoint, normal linear functionals;
\item $\alpha$ preserves the set of positive linear functionals;
\item if $\omega$ is a faithful, positive linear functional, then so is $\alpha(\gr,\,\omega)$ for every $\gr\in\gapp$.
\end{enumerate}

\proof
First of all, we note that the second and third points follow by direct inspection.

Then, concerning the first point, we recall that a normal linear functional $\xi$ is a continuous linear functional which is also continuous with respect to the weak* topology on $\appa$ generated by its topological predual $\appa_{*}$.
Equivalently, for every normal linear functional $\xi\in\appa^{*}$ there is an element $\widetilde{\xi}\in\appa_{*}$ such that $\xi=i_{**}(\widetilde{\xi})$ where $i_{**}$ is the canonical inclusion of $\appa_{*}$ in its double dual $\appa^{*}$.
Then, for every $\mathbf{b}\in\appa$, the maps
\be
\begin{split}
l_{\mathbf{b}}\,&\colon\appa\rightarrow\appa,\;\;\;\; l_{\mathbf{b}}(\mathbf{a}):=\mathbf{ba} \\
r_{\mathbf{b}}\,&\colon\appa\rightarrow\appa,\;\;\;\;  r_{\mathbf{b}}(\mathbf{a}):=\mathbf{ab}
\end{split}
\ee
are continuous with respect to the weak* topology on $\appa$ generated by its topological predual $\appa_{*}$, and it it is immediate to check that the linear functional $\alpha(\gr,\,\xi)\colon\appa\rightarrow\mathbb{C}$ may be written as
\be
\alpha(\gr,\,\xi)\,=\,\xi\,\circ\,l_{\gr^{\dagger}}\,\circ\,r_{\gr}\,,
\ee
which means that $\alpha(\gr,\,\xi)$ is weak* continuous. 
\qed
\end{proposition}

Let $\mathcal{O}_{sa}\subset\appa^{*}_{sa}$ be an orbit of $\gapp$ by means of $\alpha$.
Considering $\xi\in\mathcal{O}_{sa}$ and the coset space $\gapp/\gapp_{\xi}$, where $\gapp_{\xi}$ is the isotropy subgroup
\be\label{eqn: isotropy subgroup of linear action}
\gapp_{\xi}\,=\,\left\{\gr\in\gapp\,\colon\:\:\alpha(\gr,\,\xi)\,=\, \xi \right\} ,
\ee 
of $\xi$ with respect to $\alpha$,  the map $i_{\xi}^{\alpha}\colon \gapp/\gapp_{\xi}\rightarrow \mathcal{O}_{sa}$ given by
\be
[\gr]\mapsto i_{\xi}^{\alpha}([\gr])=\alpha(\gr,\xi)
\ee
provides a set-theoretical bijection between the coset space $\gapp/\gapp_{\xi}$   and the orbit $\mathcal{O}_{sa}$ for every $\xi\in\mathcal{O}_{sa}\subset\appa^{*}_{sa}$.
According to the results recalled in appendix \ref{app: banach-lie groups and homogeneous spaces}, this means that  we may dress the orbit $\mathcal{O}_{sa}$ with the structure of homogeneous Banach manifold of $\gapp$ whenever the isotropy subgroup $\gapp_{\xi}$ is a Banach-Lie subgroup of $\gapp$.
Specifically, it is the quotient space $\gapp/\gapp_{\xi}$ that is endowed with the structure of homogeneous Banach manifold, and this structure may be ``transported'' to $\mathcal{O}_{sa}$  in view of the bijection $ i_{\xi}^{\alpha}$ between $\gapp/\gapp_{\xi}$ and $\mathcal{O}_{sa}$.

In general, the fact that $\gapp_{\xi}$ is a Banach-Lie subgroup of $\gapp$ depends on both $\xi$ and $\appa$.
However, we will now see that $\gapp_{\xi}$ is an algebraic subgroup of $\gapp$ for every $\xi$ and every unital $C^*$-algebra $\appa$.
According to \cite[p. 117]{upmeier-symmetric_banach_manifolds_and_jordan_calgebras}, a subgroup $\mathcal{K}$ of $\gapp$ is called algebraic of order $n$ if there is a family $Q$ of Banach-space-valued continuous polynomials on $\appa\times\appa$ with degree at most $n$ such that
\be
\mathcal{K}=\left\{\gr\in\gapp\colon p(\gr\,,\gr^{-1})=0\;\;\forall p\in Q\right\}\,.
\ee

\begin{proposition}\label{prop: isotropy subgroups of linear action on positive linear functionals are algebraic subgroups}
The isotropy subgroup  $\gapp_{\xi}$ of $\xi\in\appa^{*}_{sa}$ is an algebraic subgroup of $\gapp$ of order $2$ for every $\xi\in\appa^{*}_{sa}$.

\proof
Define the family $Q_{\xi}=\{p_{\xi,\mathbf{c}}\}_{\mathbf{c}\in\appa}$ of complex-valued polynomials of order $2$ as follows\footnote{Note that the dependence of $p_{\xi,\mathbf{c}}$ on the second variable is trivial, and this explains why $\mathbf{b}$ does not appear on the rhs.}:
\be
p_{\xi,\mathbf{c}}(\mathbf{a}\,,\mathbf{b})\,:=\, \xi\left(\mathbf{c}\right) \,-\, \xi\left(\mathbf{a}^{\dagger}\,\mathbf{c}\mathbf{a}\right)\,.
\ee 
The continuity of every $p_{\xi,\mathbf{c}}$ follows easily from the fact that $\xi$ is a norm-continuous linear functional on $\appa$.
A moment of reflection shows that
\be
\gapp_{\xi}=\left\{\gr\in\gapp \colon p_{\xi,\mathbf{c}}(\gr\,,\gr^{-1})=0\;\;\forall p_{\xi,\mathbf{c}}\in Q_{\xi}\right\}\,,
\ee
and thus $\gapp_{\xi}$ is an algebraic subgroup of $\gapp$ of order $2$ for all $\xi\in\appa^{*}_{sa}$. 
\qed

\end{proposition}

Being an algebraic subgroup of $\gapp$, the isotropy subgroup $\gapp_{\xi}$ is  a closed subgroup of $\gapp$ which is also a real Banach-Lie group in the relativised norm topology, and its Lie algebra $\mathfrak{g}_{\xi}\subset\mathfrak{g}$ is given by the closed subalgebra (see  \cite[p. 667]{harris_kaup-linear_algebraic_groups_in_infinite_dimensions}, and \cite[p. 118]{upmeier-symmetric_banach_manifolds_and_jordan_calgebras})
\be
\mathfrak{g}_{\xi}=\left\{\mathbf{a}\in\mathfrak{g}\equiv\appa\,\colon \exp(t\mathbf{a})\in\gapp_{\xi}\;\forall t\in\mathbb{R}\right\}\,.
\ee
According to proposition \ref{prop: characterization of Banach-Lie subgroups}, the isotropy subgroup $\gapp_{\xi}$ is a Banach-Lie  subgroup of $\gapp$ if and only if the Lie algebra $\mathfrak{g}_{\xi}$ of $\gapp_{\xi}$ is a split subspace of $\mathfrak{g}=\appa$ and $\exp(V)$ is a neighbourhood of the identity element in $\gapp_{\xi}$ for every neighbourhood $V$ of $\mathbf{0}\in\mathfrak{g}_{\xi}$ (see \cite[p. 129]{upmeier-symmetric_banach_manifolds_and_jordan_calgebras} for an explicit proof).
The fact that $\exp(V)$ is a neighbourhood of the identity element in $\gapp_{\xi}$ for every neighbourhood $V$ of $\mathbf{0}\in\mathfrak{g}_{\xi}$ follows from the fact that $\gapp_{\xi}$ is an algebraic subgroup of $\gapp$ (see \cite[p. 667]{harris_kaup-linear_algebraic_groups_in_infinite_dimensions}).

Next, if $\mathbf{a}\in \mathfrak{g}=\appa$, we have that
\be
\gr_{t}\,=\,\exp(t\mathbf{a}) 
\ee
is a smooth curve in $\gapp$ for all $t\in\mathbb{R}$.
Consequently, we have the smooth curve $\xi_{t}$ in $\appa^{*}$ given by
\be
\xi_{t}(\mathbf{b})\,=\,(\alpha(\gr_{t},\,\xi))(\mathbf{b})\,=\,\xi\left(\gr_{t}^{\dagger}\,\mathbf{b}\,\gr_{t}\right) 
\ee
for all $t\in\mathbb{R}$ and for all $\mathbf{b}\in\appa$.
Therefore, we may compute
\be\label{eqn: fundamental vector field of linear action}
\begin{split}
\frac{\mathrm{d}}{\mathrm{d}\,t}\,\left(\xi\left(\gr_{t}^{\dagger}\,\mathbf{b}\,\gr_{t}\right)\right)_{t=0}&\,=\,\lim_{t\ra 0}\, \frac{1}{t}\,\left( \xi\left(\gr_{t}^{\dagger}\,\mathbf{b}\,\gr_{t}\right) - \xi(\mathbf{b})\right)\,=\,\\
&\,=\,\lim_{t\ra 0}\, \frac{1}{t}\,\sum_{j,k=0}^{+\infty}\,\left( \xi\left(\frac{(t\mathbf{a}^{\dagger})^{k}}{k!}\,\mathbf{b}\,\frac{(t\mathbf{a})^{j}}{j!}\right)-\xi(\mathbf{b})\right) \,=\,\\
&\,=\,\xi\left(\mathbf{a}^{\dagger}\,\mathbf{b}\,+\,\mathbf{b}\,\mathbf{a}\right)\,
\end{split}
\ee
for every $\mathbf{b}\in\appa$, from which it follows that $\mathbf{a}$ is in the Lie algebra $\mathfrak{g}_{\xi}$ of the isotropy group $\gapp_{\xi}$ if and only if
\be\label{eqn: isotropy subalgebra for linear action}
\xi\left(\mathbf{a}^{\dagger}\,\mathbf{b}\,+\,\mathbf{b}\,\mathbf{a}\right)\,=0
\ee
for every $\mathbf{b}\in\appa$.
In particular, note that the identity operator $\mathbb{I}$ never belongs to $\mathfrak{g}_{\xi}$.

When $\mathrm{dim}(\appa)=N<\infty$, the Lie algbera $\mathfrak{g}_{\xi}$ is a split subspace for every $\xi\in\appa^{*}_{sa}$, and thus every orbit of $\gapp$ in $\appa^{*}_{sa}$ by means of $\alpha$ is a homogeneous Banach manifold of $\gapp$.
Clearly, when $\appa$ is infinite-dimensional, this is no-longer true, and a case by case analysis is required.
For instance, in subsection \ref{sebsec: positive trace-class operators}, we will show that $\mathfrak{g}_{\xi}$ is a split subspace of $\mathfrak{g}=\appa$ when $\appa$ is the algebra $\bh$ of bounded linear operators on a complex, separable Hilbert space $\hh$, and $\xi$ is any normal, positive linear functional on $\bh$ (positive, trace-class linear operator on $\hh$).
This means that all the orbits of $\gapp=\mathcal{GL}(\hh)$ passing through normal, positive linear functionals are homogeneous Banach manifolds of $\gapp$, and we will classify these orbits into four different types.
Furthermore, in subsection \ref{subsec: faithful, finite trace}, we will prove that $\mathfrak{g}_{\xi}$ is a split subspace of $\mathfrak{g}=\appa$ whenever $\xi$ is a faithful, finite trace on $\appa$.

Now, \grit{suppose} $\xi$ is such that $\mathfrak{g}_{\xi}$ is a split subspace of $\appa$, that is, the isotropy subgroup $\gapp_{\xi}$ is a Banach-Lie subgroup of $\gapp$.
In this case, the orbit $\mathcal{O}_{sa}$ containing $\xi$ is endowed with a Banach manifold structure such that the map $\tau_{\xi}^{\alpha}\colon\gapp\rightarrow \mathcal{O}_{sa}$ given by
\be\label{eqn: submersion for alpha}
\gr\,\mapsto\,\tau_{\xi}^{\alpha}(\gr)\,:=\,\alpha(\gr,\xi)
\ee
is a smooth surjective submersion for every $\xi\in\mathcal{O}_{sa}$.
Moreover, $\gapp$ acts transitively and smoothly on $\mathcal{O}_{sa}$, and  the tangent space $T_{\xi}\mathcal{O}_{sa}$ at $\xi\in\mathcal{O}_{sa}$ is diffeomorphic to $\mathfrak{g}/\mathfrak{g}_{\xi}$ (see \cite[p. 105]{bourbaki-groupes_et_algebres_de_lie} and  \cite[p. 136]{upmeier-symmetric_banach_manifolds_and_jordan_calgebras}).
Note that this smooth differential structure on  $\mathcal{O}_{sa}$  is unique up to smooth diffeomorphism.
The canonical immersion $i_{sa}\colon\mathcal{O}_{sa}\lra\appa^{*}_{sa}$ given by $i_{sa}(\xi)=\xi$ for every $\xi\in\mathcal{O}_{sa}$ is easily seen to be a smooth map, and its tangent map is injective for every point in the orbit.

\begin{proposition}\label{prop: canonical immersion of the orbits through positive linear functionals is smooth}
Let $\xi$ be such that the isotropy subgroup $\gapp_{\xi}$ is a Banach-Lie subgroup of $\gapp$, let $\mathcal{O}_{sa}$ be the orbit containing $\xi$ endowed with the smooth structure coming from $\gapp$, and consider the map $l_{\mathbf{a}}\colon\,\mathcal{O}_{sa}\lra\mathbb{R}$, with $\mathbf{a}$ a self-adjoint element in $\appa$, given by
\be
l_{\mathbf{a}}(\xi)\,:=\,\xi(\mathbf{a}).
\ee
Then:
\begin{enumerate}
\item the canonical immersion map $i_{sa}\colon\mathcal{O}_{sa}\lra\appa^{*}_{sa}$ is  smooth;
\item the map  $l_{\mathbf{a}}\colon\,\mathcal{O}_{sa}\lra\mathbb{R}$ is smooth;
\item the tangent map $T_{\xi}i_{sa}$ at $\xi\in\mathcal{O}_{sa}$ is injective for all $\xi$ in the orbit.
\end{enumerate}

\proof

\begin{enumerate}
\item We will exploit proposition \ref{prop: smoothness of maps from homogeneous space is related with  smoothness of maps from group} in appendix \ref{app: banach-lie groups and homogeneous spaces} in order to prove the smoothness of the immersion $i_{sa}$.
Specifically, we consider the map
\be
\alpha_{\xi}\,\colon\,\gapp\,\lra\,\appa^{*}_{sa},\;\;\;\alpha_{\xi}(\gr)\,:=\,\alpha(\gr,\omega)
\ee
where $\alpha$ is the action of $\gapp$ on $\appa^{*}_{sa}$ defined by equation \eqref{eqn: linear action of invertible elements on the dual}, and note that, quite trivially, it holds
\be
\alpha_{\xi}\,=\,i_{sa}\,\circ\,\tau_{\xi}^{\alpha}.
\ee
Consequently, being $\tau_{\xi}^{\alpha}$ a smooth submersion for every $\xi\in\mathcal{O}$,  proposition \ref{prop: smoothness of maps from homogeneous space is related with  smoothness of maps from group} in appendix \ref{app: banach-lie groups and homogeneous spaces} implies that the immersion $i_{sa}$ is smooth if $\alpha_{\xi}^{\alpha}$ is smooth.
Clearly, $\alpha_{\xi}$ is smooth because $\alpha$ is a smooth action according to proposition \ref{prop: the map alpha is smooth} and the discussion below.

\item Regarding the second point, it suffices to note that $l_{\mathbf{a}}$ is the composition of the linear (and thus smooth) map $L_{\mathbf{a}}\colon\appa^{*}_{sa}\lra\mathbb{R}$ given by 
\be\label{eqn: linear functions associated with elements in the predual}
L_{\mathbf{a}}(\xi)\,=\, \xi(\mathbf{a})
\ee
with the canonical immersion $i_{sa}\colon\mathcal{O}_{sa}\lra\appa^{*}_{sa}$ which is smooth because of what has been proved above.

\item Now, consider the family $\{l_{\mathbf{a}}\}_{\mathbf{a}\in\appa}$ of smooth functions on the orbit $\mathcal{O}_{sa}$, and suppose that $\mathbf{V}_{\xi}$ and $\mathbf{W}_{\xi}$ are tangent vectors at $\xi\in\mathcal{O}_{sa}$  such that
\be\label{eqn: equation for tangent vectors, positive linear functionals}
\langle (\mathrm{d}l_{\mathbf{a}})_{\xi};\,\mathbf{V}_{\xi}\rangle\,=\,\langle (\mathrm{d}l_{\mathbf{a}})_{\xi};\,\mathbf{W}_{\xi}\rangle
\ee
for every $\mathbf{a}\in\appa_{sa}$.
Since $l_{\mathbf{a}}=L_{\mathbf{a}}\circ\,i_{sa}$, we have
\be
\langle (\mathrm{d}l_{\mathbf{a}})_{\xi};\,\mathbf{V}_{\xi}\rangle\,=\,\langle (\mathrm{d}L_{\mathbf{a}})_{i_{sa}(\xi)};\,T_{\xi}i_{sa}(\mathbf{V}_{\xi})\rangle
\ee
and
\be
\langle (\mathrm{d}l_{\mathbf{a}})_{\xi};\,\mathbf{W}_{\xi}\rangle\,=\,\langle (\mathrm{d}L_{\mathbf{a}})_{i_{sa}(\xi)};\,T_{\xi}i(\mathbf{W}_{\xi})\rangle
\ee
Note that the family of linear functions of the type $L_{\mathbf{a}}$ with  $\mathbf{a}\in\appa_{sa}$ (see equation \eqref{eqn: linear functions associated with elements in the predual}) are enough to separate the tangent vectors at $\xi$ for every $\xi\in\appa^{*}_{sa}$ because the tangent space at $\xi\in\appa^{*}_{sa}$ is diffeomorphic with $\appa^{*}_{sa}$ in such a way that 
\be
\langle (\mathrm{d}L_{\mathbf{a}})_{\xi};\mathbf{V}_{\xi}\rangle\,=\,\mathbf{V}_{\xi}(\mathbf{a})\,=\,L_{\mathbf{a}}(\mathbf{V}_{\xi})
\ee
for every $\mathbf{V}_{\xi}\in T_{\xi}\appa^{*}_{sa}\cong\appa^{*}_{sa}$, 
and $\appa_{sa}$ (the predual of $\appa^{*}_{sa}$) separates the points of $\appa^{*}_{sa}$ (see \cite{kaijser-a_note_on_dual_banach_spaces}).
Consequently, since $T_{\xi}i_{sa}(\mathbf{V}_{\xi})$ and $T_{\xi}i_{sa}(\mathbf{W}_{\xi})$ are tangent vectors at $i_{sa}(\xi)\in\appa^{*}_{sa}$ and the functions $L_{\mathbf{a}}$ with $\mathbf{a}\in\appa_{sa}$ are enough to separate them and we conclude that the validity of equation \eqref{eqn: equation for tangent vectors, positive linear functionals} for all $\mathbf{a}\in\appa_{sa}$ is equivalent to 
\be\label{egn; tangent map of orbit immersion, linear action}
T_{\xi}i_{sa}(\mathbf{V}_{\xi})\,=\,T_{\xi}i_{sa}(\mathbf{W}_{\xi}).
\ee
Then, if $\gr_{t}\,=\,\exp(t\mathbf{a})$ is a one-parameter subgroup in $\gapp$ so that
\be
\xi_{t}\,=\,\alpha(\gr_{t},\,\xi)
\ee
is a smooth curve in $\mathcal{O}$ starting at $\xi$ with associated tangent vector $\mathbf{V}_{\xi}$,   we have
\be
\langle (\mathrm{d}L_{\mathbf{b}})_{i_{sa}(\xi)};\,T_{\xi}i_{sa}(\mathbf{V}_{\xi})\rangle\,=\,\frac{\mathrm{d}}{\mathrm{d}t}\,\left(L_{\mathbf{b}}\,\circ\,i_{sa}(\xi_{t})\right)_{t=0}
\ee
which we may compute in analogy with equation \eqref{eqn: fundamental vector field of linear action} obtaining
\be\label{eqn: tangent vector at a point in the orbit of alpha}
\begin{split}
\frac{\mathrm{d}}{\mathrm{d}t}\,\left(L_{\mathbf{b}}\,\circ\,i_{sa}(\xi_{t})\right)_{t=0}&\,=\,\xi\left(\mathbf{a}^{\dagger}\,\mathbf{b}\,+\,\mathbf{b}\,\mathbf{a}\right).
\end{split}
\ee
Comparing equation \eqref{eqn: tangent vector at a point in the orbit of alpha} with equation \eqref{eqn: isotropy subalgebra for linear action} we conclude that $\mathbf{V}_{\xi}$ and $\mathbf{W}_{\xi}$ satisfy equation \eqref{egn; tangent map of orbit immersion, linear action} if and only if they coincide, and thus $T_{\xi}i_{sa}$ is injective for all $\xi\in\mathcal{O}_{sa}$.
\end{enumerate}

\qed
 
\end{proposition}

It is important to note that the topology  underlying the differential structure on the orbit $\mathcal{O}_{sa}$ containing $\xi$ comes from the topology of $\gapp$ in the sense that a subset $U$ of the orbit is open iff $(\tau^{\alpha}_{\xi})^{-1}(U)$ is open in $\gapp$.
In principle, this topology on $\mathcal{O}_{sa}$ has nothing to do with the topology of $\mathcal{O}_{sa}$ when thought of as a subset of $\appa^{*}_{sa}$ endowed with the relativised norm topology, or with the relativised weak* topology.
However, from proposition \ref{prop: canonical immersion of the orbits through positive linear functionals is smooth}, it follows that the map $l_{\mathbf{a}}\colon\,\mathcal{O}_{sa}\lra\mathbb{R}$ is continuous for every $\mathbf{a}\in\appa_{sa}$, and we may conclude that the topology underlying the homogeneous Banach manifold structure on $\mathcal{O}_{sa}$ is stronger than the relativised weak* topology coming from $\appa^{*}_{sa}$.

In general, the action $\alpha$ does not preserve the space of states $\stsp$ on $\appa$.
At this purpose, in section \ref{sec: state-preserving action}, we provide a modification of $\alpha$ that allows us to overcome this situation.

\subsection{Positive, trace-class operators}\label{sebsec: positive trace-class operators}

Let $\hh$ be a complex, separable Hilbert space and denote by $\appa$ the $W^{*}$-algebra $\bh$ of bounded, linear operators on $\hh$.
The predual of $\appa$ may be identified with the space $\mathcal{T}(\hh)$ of trace-class linear operators on $\hh$ (see \cite[p. 61]{takesaki-theory_of_operator_algebra_I}).
In particular, a normal, self-adjoint linear functional $\widetilde{\xi}$ on $\appa$ may be identified with a self-adjoint, trace-class operator $\xi$ on $\hh$, and the duality relation may be expressed by means of the trace operation
\be
\widetilde{\xi}(\mathbf{a})\,=\,\mathrm{Tr}\,\left(\xi\,\mathbf{a}\right)
\ee
for all $\mathbf{a}\in\appa=\mathcal{B}(\hh)$.
Furthermore, it is known that $\appa=\mathcal{B}(\hh)$ may be identified with the double dual of the $C^{*}$-algebra $\mathcal{K}(\hh)$ of compact, linear operators on $\hh$ in such a way that the linear functionals on $\mathcal{K}(\hh)$ are identified with the normal linear functionals on $\appa$ (see \cite[p. 64]{takesaki-theory_of_operator_algebra_I}).

Now, we will study the orbits of the action $\alpha$ of the group $\gapp$ of invertible elements in $\appa$ on the normal, positive linear functionals on $\appa$.
The group $\gapp$ is the Banach-Lie group $\mathcal{GL}(\hh)$ of invertible, bounded linear operators on the complex,  separable Hilbert space  $\mathcal{H}$, and its action $\alpha$ on a self-adjoint, normal linear functional $\widetilde{\xi}$ reads 
\be\label{eqn: linear action on self-adjoint trace-class}
\left(\alpha(\gr,\,\widetilde{\xi})\right)(\mathbf{a})\,=\,\widetilde{\xi}\left(\gr^{\dagger}\,\mathbf{a}\,\gr\right)\,=\,\mathrm{Tr}\left(\xi\,\gr^{\dagger}\,\mathbf{a}\,\gr\right)\;\;\;\forall\,\mathbf{a}\in\appa=\bh  .
\ee
Equivalently, we may say that $\alpha$ transform the element $\xi$ in the predual $(\appa_{*})_{sa}=(\mathcal{T}(\hh))_{sa}$ of $\appa_{sa}$  in the element $\xi_{\gr}$ given by
\be\label{eqn: linear action on self-adjoint trace-class 2}
\xi_{\gr}\,=\,\gr\,\xi\,\gr^{\dagger}\,.
\ee
This last expression allows us to work directly with trace-class operators.

According to the spectral theory for compact operators (see \cite[ch. VII]{reed_simon-methods_of_modern_mathematical_physics_I_functional_analysis}), given a positive, trace-class linear operator $\varrho\,\neq\,\mathbf{0}$ on $\hh$, there is a decomposition 
\be\label{eqn: spectral decomposition of positive trace-class operator}
\mathcal{H}=\mathcal{H}_{\varrho}\oplus\mathcal{H}_{\varrho}^{\perp}
\ee
and a countable orthonormal basis $\{|e_{j}\rangle,\,|f_{j}\rangle\}$ adapted to this decomposition such that $\varrho$ can be written as
\be\label{eqn: spectral decomposition of positive trace-class operator 2}
\varrho\,=\,\sum_{j=1}^{\mathrm{dim}(\mathcal{H}_{\varrho})}\, p^{j}\,|e_{j}\rangle\langle e_{j}|\,,
\ee
with $\mathrm{dim}(\mathcal{H}_{\varrho})>0$ and  $p^{j}>0$ for all $j\in[1,...,\mathrm{dim}(\mathcal{H}_{\varrho})]$.
In general, we have four different situations:

\begin{enumerate}
\item $0<\mathrm{dim}(\mathcal{H}_{\varrho})=N<\infty$;
\item $\mathrm{dim}(\mathcal{H}_{\varrho})=\infty$ and $0<\mathrm{dim}(\mathcal{H}^{\perp}_{\varrho})=M<\infty$;
\item $\mathrm{dim}(\mathcal{H}_{\varrho})=\infty$ and $\mathrm{dim}(\mathcal{H}^{\perp}_{\varrho})=0$
\item $\mathrm{dim}(\mathcal{H}_{\varrho})=\mathrm{dim}(\mathcal{H}_{\varrho}^{\perp})=\infty$,
\end{enumerate}
and we set
\be\label{eqn: partition of positive trace-class operators}
\begin{split}
\pposn_{N}&\,:=\,\left\{\mathbf{0}\,\neq\,\varrho\in\posn\,\,|\;0<\mathrm{dim}(\mathcal{H}_{\varrho})=N<\infty\right\} \\ 
\pposn_{M}^{\perp}&\,:=\,\left\{\mathbf{0}\,\neq\,\varrho\in\posn\,\,|\;\mathrm{dim}(\mathcal{H}_{\varrho})=\infty\,\mbox{ and }\,0<\mathrm{dim}(\mathcal{H}^{\perp}_{\varrho})=M<\infty\right\} \\ 
\pposn_{0}^{\perp}&\,:=\,\left\{\mathbf{0}\,\neq\,\varrho\in\posn\,\,|\;\mathrm{dim}(\mathcal{H}_{\varrho})=\infty\,\mbox{ and }\,\mathrm{dim}(\mathcal{H}^{\perp}_{\varrho})=0\right\} \\ 
\pposn_{\infty}&\,:=\,\left\{\mathbf{0}\,\neq\,\varrho\in\posn\,\,|\;\mathrm{dim}(\mathcal{H}_{\varrho})=\mathrm{dim}(\mathcal{H}_{\varrho}^{\perp})=\infty\right\} .
\end{split}
\ee
The subscripts here denote either the dimension of the space on which $\varrho$  operates, or its codimension when the symbol $\perp$ is used.
Clearly, when $\mathrm{dim}(\hh)<\infty$, we have $\pposn_{N}=\emptyset$ for all $N>\mathrm{dim}(\hh)$, and $\pposn_{M}^{\perp}=\pposn_{0}^{\perp}=\pposn_{\infty}=\emptyset$.

The advantage of working with a separable Hilbert space is that every bounded linear operator $\mathbf{a}\in\bh=\appa$ may be looked at as an infinite matrix whose matrix elements $a_{jk}$ are given by
\be
a_{jk}\,=\,\langle e_{j}|\mathbf{a}|e_{k}\rangle
\ee
where $\{|e_{j}\rangle\}$ is an orthonormal basis in $\hh$.
Clearly, the matrix describing $\mathbf{a}$ depends on the choice of the orthonormal basis.
However, once this choice is made, we may translate the algebraic operations in $\bh=\appa$, like the sum, the multiplication, and the involution, in the language of matrix algebras (see \cite[p. 48]{akhiezer_glazman-theory_of_linear_operators_in_hilbert_space}).

In particular, if $\varrho\,\neq\,\mathbf{0}$ is a positive, trace-class linear operator, we may choose a  countable orthonormal basis $\{|e_{j}\rangle,\,|f_{j}\rangle\}$ adapted to the spectral decomposition of $\varrho$ so that the matrix associated with $\varrho$ is diagonal.
On the other hand, the matrix expression $A$ of $\mathbf{a}\in\appa=\bh$ with respect to the countable orthonormal basis $\{|e_{j}\rangle,\,|f_{j}\rangle\}$ adapted to the spectral decomposition of $\varrho$ reads
\be
A\,=\,
\left(\begin{matrix} 
A_{1} & A_{2} \\
A_{3} & A_{4}
\end{matrix}\right),
\ee
where $A_{1}$ may be thought of as a bounded linear operator sending $\mathcal{H}_{\varrho}$ in itself, $A_{2}$  may be thought of as a bounded linear operator sending  $\hh_{\varrho}^{\perp}$ in $\mathcal{H}_{\varrho}$, $A_{3}$  may be thought of as a bounded linear operator sending  $\mathcal{H}_{\varrho}$  in $\mathcal{H}_{\varrho}^{\perp}$, and $A_{4}$  may be thought of as a bounded linear operator sending  $\mathcal{H}_{\varrho}^{\perp}$ in itself.

\begin{proposition}\label{prop: characterization of isotropy subalgebra of a positive, trace-class operator for the linear action}
Let $\hh$ be a complex, separable Hilbert space, let $\varrho$ be a positive, trace-class linear operator on $\hh$, and denote by $\{|e_{k}\rangle,\,|f_{l}\rangle\}$ the orthonormal basis of $\hh$ adapted to the spectral decomposition of $\varrho$ (see equations \eqref{eqn: spectral decomposition of positive trace-class operator} and \eqref{eqn: spectral decomposition of positive trace-class operator 2}).
Then, the Lie algebra $\mathfrak{g}_{\varrho}$ of the isotropy subgroup $\gapp_{\varrho}$ of $\varrho$ with respect to the action $\alpha$ in equations \eqref{eqn: linear action on self-adjoint trace-class} and \eqref{eqn: linear action on self-adjoint trace-class 2} is given by
\be
\mathfrak{g}_{\varrho}\,=\,\left\{\mathbf{a}\in\appa\,\,\colon \footnotesize{\begin{matrix}\,\langle f_{k}|\mathbf{a}|f_{l}\rangle\,\mbox{ arbitrary } \,\,\forall\,k,l\in[1,...,\mathrm{dim}(\hh_{\varrho}^{\perp})];\\ \\
 \,\langle e_{k}|\mathbf{a}|f_{l}\rangle\,\mbox{ arbitrary } \,\,\forall\,l\in[1,...,\mathrm{dim}(\hh_{\varrho}^{\perp})],\,\forall\,k\in[1,...,\mathrm{dim}(\hh_{\varrho})] \\ \\
 \,\langle f_{l}|\mathbf{a}|e_{k}\rangle\,=\,0 \,\,\forall\,l\in[1,...,\mathrm{dim}(\hh_{\varrho}^{\perp})],\,\forall\,k\in[1,...,\mathrm{dim}(\hh_{\varrho})] \\ \\
\langle e_{k}|\mathbf{a}|e_{l}\rangle\,=\, - \frac{p_{k}}{p_{l}}\,\overline{\langle e_{l}|\mathbf{a}|e_{k}\rangle}\,\, \,\,\forall\,k,l\in[1,...,\mathrm{dim}(\hh_{\varrho})]  \end{matrix}} \right\}.
\ee

\proof
Recall that an element $\mathbf{a}\in\mathfrak{g}=\appa=\bh$ is in the Lie algebra $\mathfrak{g}_{\varrho}$ of the isotropy subgroup $\gapp_{\varrho}$ of $\varrho$ if and only if (se equation \eqref{eqn: isotropy subalgebra for linear action})
\be\label{eqn: isotropy subalgebra for positive trace-class 1}
\widetilde{\varrho}(\mathbf{a}^{\dagger}\,\mathbf{b} + \mathbf{b}\,\mathbf{a})\,=\,
\mathrm{Tr}\left(\varrho\,(\mathbf{a}^{\dagger}\,\mathbf{b} + \mathbf{b}\,\mathbf{a})\right)\,=\,0\;\;\;\forall\,\,\,\mathbf{b}\in\appa=\bh .
\ee
Using the matrix expressions of $\varrho,\,\mathbf{a},$ and $\mathbf{b}$, a direct computation shows that equation \eqref{eqn: isotropy subalgebra for positive trace-class 1}  poses no constraints on the factor $A_{2}$ in the matrix expression of $\mathbf{a}$, or, equivalenty, we have that 
\be
\langle e_{k}|\mathbf{a}|f_{l}\rangle\,\mbox{ is arbitrary } \,\,\forall\,l\in[1,...,\mathrm{dim}(\hh_{\varrho}^{\perp})],\,\forall\,k\in[1,...,\mathrm{dim}(\hh_{\varrho})].
\ee
Then, since $\mathbf{b}$ in equation \eqref{eqn: isotropy subalgebra for positive trace-class 1} is arbitrary, if we fix $k\in[1,...,\mathrm{dim}(\hh_{\varrho})]$ and $l\in[1,...,\mathrm{dim}(\hh_{\varrho}^{\perp})]$ and take $\mathbf{b}=|e_{k}\rangle\langle f_{l}|$, equation \eqref{eqn: isotropy subalgebra for positive trace-class 1} becomes
\be
p^{k}\,\langle f_{l}|\mathbf{a}| e_{k}\rangle\,=\,0\,\,\Longleftrightarrow\,\,\langle f_{l}|\mathbf{a}| e_{k}\rangle\,=\,0.
\ee
Clearly, we may do this for every $k\in[1,...,\mathrm{dim}(\hh_{\varrho})]$ and for every $l\in[1,...,\mathrm{dim}(\hh_{\varrho}^{\perp})]$, which means that if $\mathbf{a}$ is in the isotropy algebra $\mathfrak{g}_{\varrho}$, then $A_{3}=0$.
Similarly, if we fix $k\in[1,...,\mathrm{dim}(\hh_{\varrho})]$ and $l\in[1,...,\mathrm{dim}(\hh_{\varrho}^{\perp})$ and take $\mathbf{b}\,=\,|f_{l}\rangle\langle e_{k}|$, equation \eqref{eqn: isotropy subalgebra for positive trace-class 1} becomes
\be
p^{k}\,\overline{\langle f_{l}|\mathbf{a}| e_{k}\rangle}\,=\,0\,
\ee
which is equivalent to the previous equation.
Then, if we fix $k\in[1,...,\mathrm{dim}(\hh_{\varrho}^{\perp})$ and take $\mathbf{b}=|f_{k}\rangle\langle f_{l}|$, we immediately see that equation \eqref{eqn: isotropy subalgebra for positive trace-class 1} poses no constraints on $\mathbf{a}$.
Putted differently, if $\mathbf{a}$ is in $\mathfrak{g}_{\varrho}$, then the factor $A_{4}$ in the matrix expression of $\mathbf{a}$ is arbitrary.
Next, we take  $\mathbf{b}=|e_{l}\rangle\langle e_{k}|$ with $l,k\in[1,...,\mathrm{dim}(\hh_{\varrho})]$, and a direct computation shows that we must have
\be\label{eqn: isotropy algebra on the eigenspace}
\langle e_{k}|\mathbf{a}|e_{l}\rangle\,=\, - \frac{p_{k}}{p_{l}}\,\overline{\langle e_{l}|\mathbf{a}|e_{k}\rangle}\,.
\ee
Consequently, noting that every $\mathbf{a}\in\appa$ may be written as the sum of two self-adjoint elements in $\appa$, say $\mathbf{x}$ and $\mathbf{y}$, as follows
\be
\mathbf{a}\,=\,\mathbf{x} + \,\imath\,\mathbf{y} ,
\ee
we immediately obtain that equation \eqref{eqn: isotropy algebra on the eigenspace} is equivalent to 
\be
x_{kl}^{e}\,=\,\imath\,\frac{p_{k}-p_{l}}{p_{k}+p_{l}}\,y_{kl}^{e},
\ee
where 
\be
x_{kl}^{e}\,=\,\langle e_{k}|\mathbf{x}| e_{l}\rangle ,\;\;\;y_{kl}^{e}\,=\,\langle e_{k}|\mathbf{y}| e_{l}\rangle .
\ee
This means that the self-adjoint part of $\mathbf{a}$ on $\hh_{\rho}$ is uniquely determined by the (arbitrary) skew-adjoint part of $\mathbf{a}$ on $\hh_{\rho}$ unless we are considering a subspace where $\rho$ acts as a multiple of the identity, in which case, the self-adjoint part identically vanishes.

\qed

 \end{proposition}

The characterization of $\mathfrak{g}_{\varrho}$ given in proposition \ref{prop: characterization of isotropy subalgebra of a positive, trace-class operator for the linear action} may be aesthetically unpleasant, but it allows to find immediately an algebraic complement for $\mathfrak{g}_{\varrho}$ in $\mathfrak{g}=\appa=\bh$.
Indeed, if we set

\be\label{eqn: complement of the isotropy subalgebra of a positive trace-class operator for the linear action}
\mathfrak{k}_{\varrho}\,=\,\left\{\mathbf{b}\in\appa\,\,\colon \footnotesize{ \begin{matrix}\,\langle f_{k}|\mathbf{b}|f_{l}\rangle\,=\,0 \,\,\forall\,k,l\in[1,...,\mathrm{dim}(\hh_{\varrho}^{\perp})];\\ \\
 \,\langle e_{k}|\mathbf{b}|f_{l}\rangle\,=\,0 \,\,\forall\,l\in[1,...,\mathrm{dim}(\hh_{\varrho}^{\perp})],\,\forall\,k\in[1,...,\mathrm{dim}(\hh_{\varrho})] \\ \\
 \,\langle f_{l}|\mathbf{b}|e_{k}\rangle\,=\,\mbox{ arbitrary } \,\,\forall\,l\in[1,...,\mathrm{dim}(\hh_{\varrho}^{\perp})],\,\forall\,k\in[1,...,\mathrm{dim}(\hh_{\varrho})] \\ \\

\langle e_{k}|\mathbf{b}|e_{l}\rangle\,=\,\overline{\langle e_{l}|\mathbf{b}|e_{k}\rangle} \,\,\forall\,k,l\in[1,...,\mathrm{dim}(\hh_{\varrho})]   \end{matrix}}\right\},
\ee
it is clear that $\mathfrak{g}_{\varrho}\cap\mathfrak{k}_{\varrho}\,=\,\{\mathbf{0}\}$.
Furthermore, since an arbitrary $\mathbf{c}\in\mathfrak{g}=\appa=\bh$ is uniquely determined by its matrix elements with respect to the orthonormal basis $\{|e_{k}\rangle,\,|f_{l}\rangle\}$  of $\hh$ adapted to the spectral decomposition of $\varrho$ (see equations \eqref{eqn: spectral decomposition of positive trace-class operator} and \eqref{eqn: spectral decomposition of positive trace-class operator 2}), a direct computation shows that 
\be
\mathfrak{g}\,=\,\mathfrak{g}_{\varrho}\,\oplus\,\mathfrak{k}_{\varrho},
\ee
algebraically.
Then, according to proposition \ref{prop: characterization of Banach-Lie subgroups},  we have that the orbit of $\gapp$ passing through $\varrho$ inherits a Banach manifold structure from $\gapp/\gapp_{\varrho}$ whenever $\mathfrak{k}_{\varrho}$ is closed in $\mathfrak{g}=\appa=\bh$.
The closedness of $\mathfrak{k}_{\varrho}$  in $\mathfrak{g}=\appa=\bh$ is the content of the next proposition.

\begin{proposition}\label{prop: the isotropy subalgebra of a positive trace-class operator for the linear action is complemented}
The linear subspace $\mathfrak{k}_{\varrho}\subset\mathfrak{g}=\appa=\bh$ is closed.

\proof
Let $\{\mathbf{b}_{n}\}_{n\in\mathbb{N}}$ be a sequence in $\mathfrak{k}_{\varrho}$ norm-converging to $\mathbf{b}\in\mathfrak{g}=\appa=\bh$.
The proof of this proposition reduces to a routine check of the matrix elements of $\mathbf{b}$ with respect to the orthonormal basis $\{|e_{k}\rangle,\,|f_{l}\rangle\}$  of $\hh$ adapted to the spectral decomposition of $\varrho$ in order to show that they satisfy all the conditions in equation \eqref{eqn: complement of the isotropy subalgebra of a positive trace-class operator for the linear action}.

The norm convergence of $\{\mathbf{b}_{n}\}_{n\in\mathbb{N}}$ to $\mathbf{b}\in\mathfrak{g}=\appa=\bh$ implies the convergence of the sequence $\{(F_{\psi\phi})_{n}\}_{n\in\mathbb{N}}$ with
\be
(F_{\psi\phi})_{n}\,=\,\langle\psi|\mathbf{b}_{n}|\phi\rangle
\ee
to 
\be
F_{\psi\phi}\,=\,\langle\psi|\mathbf{b}|\phi\rangle
\ee
for all $|\psi\rangle,\,|\phi\rangle\in\hh$.
In particular, if we take $|\psi\rangle=|f_{k}\rangle$ and $|\phi\rangle=|f_{l}\rangle$, with arbitrary $k,l\in[1,...,\mathrm{dim}(\hh_{\varrho}^{\perp})]$, we have
\be
0\,=\,(F_{f_{k}f_{l}})_{n}\,\stackrel{n\ra\infty}{\lra}\,\,F_{f_{k}f_{l}}\,=\,\langle f_{k}|\mathbf{b}|f_{l}\rangle
\ee
which means 
\be\label{eqn: complement is closed}
F_{f_{k}f_{l}}\,=\,\langle f_{k}|\mathbf{b}|f_{l}\rangle\,=\,0
\ee
for all $k,l\in[1,...,\mathrm{dim}(\hh_{\varrho}^{\perp})]$.
Similarly, if we take $|\psi\rangle=|e_{k}\rangle$ and $|\phi\rangle=|f_{l}\rangle$, with arbitrary $k\in[1,...,\mathrm{dim}(\hh_{\varrho})]$ and $l\in[1,...,\mathrm{dim}(\hh_{\varrho}^{\perp})]$, we obtain
\be\label{eqn: complement is closed 2}
F_{e_{k}f_{l}}\,=\,\langle e_{k}|\mathbf{b}|f_{l}\rangle\,=\,0
\ee
for all $k\in[1,...,\mathrm{dim}(\hh_{\varrho})]$ and $l\in[1,...,\mathrm{dim}(\hh_{\varrho}^{\perp})]$.

\vsp

Next, if we take $|\psi\rangle=|f_{l}\rangle$ and $|\phi\rangle=|e_{k}\rangle$, with arbitrary $k\in[1,...,\mathrm{dim}(\hh_{\varrho})]$ and $l\in[1,...,\mathrm{dim}(\hh_{\varrho}^{\perp})]$, we have that $(F_{f_{l}e_{k}})_{n}$ converges to the complex number
\be\label{eqn: complement is closed 3}
F_{f_{l}e_{k}}\,=\,\langle f_{l}|\mathbf{b}|e_{k}\rangle
\ee
and there are no constraints on $F_{f_{l}e_{k}}$ for all $k\in[1,...,\mathrm{dim}(\hh_{\varrho})]$ and $l\in[1,...,\mathrm{dim}(\hh_{\varrho}^{\perp})]$.

\vsp

Eventually,  if we take $|\psi\rangle=|e_{k}\rangle$ and $|\phi\rangle=|e_{l}\rangle$, with $k,l\in[1,...,\mathrm{dim}(\hh_{\varrho})]$, we have that $(F_{e_{k}e_{l}})_{n}$ converges to the complex number
\be
F_{e_{k}e_{l}}\,=\,\langle e_{k}|\mathbf{b}|e_{l}\rangle .
\ee
Now, since in $\mathbb{C}$ the complex conjugate of the limit is equal to the limit of the complex conjugate, we have
\be
\overline{F_{e_{k}e_{l}}}\,=\,\overline{\lim_{n\ra\infty}\,(F_{e_{k}e_{l}})_{n}}\,=\,\lim_{n\ra\infty}\,\overline{(F_{e_{k}e_{l}})_{n}}\,=\,\lim_{n\ra\infty}\,(F_{e_{l}e_{k}})_{n}\,=\,F_{e_{l}e_{k}}
\ee
which means
\be\label{eqn: complement is closed 5}
\overline{\langle e_{k}|\mathbf{b}|e_{l}\rangle}\,=\,\langle e_{l}|\mathbf{b}|e_{k}\rangle.
\ee
Comparing equations \eqref{eqn: complement is closed}, \eqref{eqn: complement is closed 2}, \eqref{eqn: complement is closed 3},  and \eqref{eqn: complement is closed 5} with the characterization of elements in $\mathfrak{k}_{\varrho}$ as given in equation \eqref{eqn: complement of the isotropy subalgebra of a positive trace-class operator for the linear action}, it follows that $\mathbf{b}$ is in $\mathfrak{k}_{\varrho}$, and thus $\mathfrak{k}_{\varrho}$ is closed in $\mathfrak{g}=\appa=\bh$.
\qed

\end{proposition}

From this it follows that, for every positive, trace-class linear operator $\varrho$ on $\hh$, the orbit $\mathcal{O}_{+}$ containing $\varrho$ (see equations \eqref{eqn: linear action on self-adjoint trace-class} and \eqref{eqn: linear action on self-adjoint trace-class 2}) is a homogeneous Banach manifold for the group $\gapp=\mathcal{GL}(\hh)$ of bounded, invertible linear operators on $\hh$.
We decided to denote by $\mathcal{O}_{+}$ the orbit containing $\varrho$ in order to emphasize the fact that elements in $\mathcal{O}$ are normal, \grit{positive} linear functionals.
Indeed, the proofs of the propositions \ref{prop: characterization of isotropy subalgebra of a positive, trace-class operator for the linear action} and \ref{prop: the isotropy subalgebra of a positive trace-class operator for the linear action is complemented} depend crucially on the positivity of $\varrho$.

We will now give a partial characterization of the orbits of $\gapp$ through normal, positive linear functionals. 
\begin{proposition}\label{prop: partial characterization of orbits of positive trace-class operators}
Let $\varrho_{0}$ and $\varrho_{1}$ be positive, trace-class operators on $\hh$, and denote by $p_{0}^{j}$ and $p_{1}^{j}$ the $j$-th eigenvalue of $\varrho_{0}$ and $\varrho_{1}$, respectively.
If  $\hh_{\varrho_{0}}$ is isomorphic to $\hh_{\varrho_{1}}$, and $\hh_{\varrho_{0}}^{\perp}$ is isomorphic to $\hh_{\varrho_{1}}^{\perp}$, and  the following condition holds
\be \label{eqn: important condition}
\frac{p_{1}^{j}}{p_{0}^{j}}\,\leq\, C \,<\,\infty\;\;\;\;\;\forall j=1,...,N,
\ee
then, the element $\gr$ defined by 
\be\label{eqn: invertible element 2}
\gr\,:=\,\sum_{j=1}^{N}\,\sqrt{\frac{p^{j}_{1}}{p^{j}_{0}}}\,|e_{j}^{1}\rangle\langle e_{j}^{0}| + \sum_{k=1}^{M}\,|f_{k}^{1}\rangle\langle f_{k}^{0}|.
\ee 
is a bounded, invertible operator on $\hh$ such that
\be\label{eqn: group relations between positive operators}
\varrho_{1}\,=\,\gr\,\varrho_{0}\,\gr^{\dagger}\,.
\ee

\proof
Since $\hh_{\varrho_{0}}$ is isomorphic to $\hh_{\varrho_{1}}$, and $\hh_{\varrho_{0}}^{\perp}$ is isomorphic to $\hh_{\varrho_{1}}^{\perp}$, the element
\be 
\gr\,:=\,\sum_{j=1}^{N}\,\sqrt{\frac{p^{j}_{1}}{p^{j}_{0}}}\,|e_{j}^{1}\rangle\langle e_{j}^{0}| + \sum_{k=1}^{M}\,|f_{k}^{1}\rangle\langle f_{k}^{0}|,
\ee
where $N=\mathrm{dim}(\hh_{\varrho_{0}})=\mathrm{dim}(\hh_{\varrho_{1}})$ and $M=\hh_{\varrho_{0}}^{\perp}=\hh_{\varrho_{1}}^{\perp}$, is well-defined, and a direct (formal) computation shows that equation \eqref{eqn: group relations between positive operators} actually holds.
All that is left to do is to check that $\gr$ is bounded and invertible.
At this purpose, we consider an arbitrary element $\psi\in\hh$ which can be written as
\be
\begin{split}
|\psi\rangle&\,=\,\sum_{j=1}^{N}\,\psi^{j}_{0}\,|e_{j}^{0}\rangle  + \sum_{k=1}^{M}\,\psi^{k}_{0,\perp}\,|f_{k}^{0}\rangle\,=\, \\
&\,=\,\sum_{j=1}^{N}\,\psi^{j}_{1}\,|e_{j}^{1}\rangle  + \sum_{k=1}^{M}\,\psi^{k}_{1,\perp}\,|f_{k}^{1}\rangle\,,
\end{split}
\ee
with respect to the orthonormal bases in $\hh$ adapted to the spectral decompositions of $\varrho_{0}$ and $\varrho_{1}$.
From this, we have
\be\label{eqn: gr is bounded}
\begin{split}
||\gr(\psi)||^{2}&\,=\,\langle\psi|\gr^{\dagger}\,\gr|\psi\rangle\,=\,\sum_{j=1}^{N}\,|\psi^{j}_{0}|^{2}\,\frac{p^{j}_{1}}{p^{j}_{0}} + \sum_{k=1}^{M}\,|\psi^{k}_{0,\perp}|^{2}\,\leq\, \\
&\,\leq\,C\sum_{j=1}^{N}\,|\psi^{j}_{0}|^{2}\, + \sum_{k=1}^{M}\,|\psi^{k}_{0,\perp}|^{2}\,\leq\, \left(C+1\right)\,||\psi||^{2}\,<\,\infty\,,
\end{split}
\ee
where we used equation \eqref{eqn: important condition} in the second passage.
Clearly, equation \eqref{eqn: gr is bounded} implies that $\gr$ is bounded.
Then,  setting
\be
\gr^{-1}\,:=\,\sum_{j=1}^{N}\,\sqrt{\frac{p^{j}_{0}}{p^{j}_{1}}}\,|e_{j}^{0}\rangle\langle e_{j}^{1}| + \sum_{k=1}^{M}\,\,|f_{k}^{0}\rangle\langle f_{k}^{1}|,
\ee
we may proceed as we did for $\gr$ to show that $\gr^{-1}$ is bounded, and a direct computation shows that $\gr^{-1}$ is the inverse of $\gr$.

\qed

\end{proposition}

Clearly, the assumptions in proposition \ref{prop: partial characterization of orbits of positive trace-class operators} are always satisfied if $\varrho_{0}$ and $\varrho_{1}$ are finite-rank operators with the same rank.

\vsp

The last step we want to take is to write down  a tangent vector $\mathbf{V}_{\varrho}$ at $\varrho\in\mathcal{O}_{+}$, where $\mathcal{O}_{+}$ is any of the orbits of $\gapp$ inside the positive, trace-class operators.
At this purpose, we consider the canonical immersion $i_{+}\colon\mathcal{O}_{+}\lra\appa^{*}_{sa}$, and we recall equation \eqref{eqn: tangent vector at a point in the orbit of alpha}, from which it follows that 
\be
T_{\varrho}i_{+}(\mathbf{V}_{\rho})(\mathbf{b})\,=\,\mathrm{Tr}\left(\varrho\,\left(\mathbf{a}^{\dagger}\,\mathbf{b} + \mathbf{b}\,\mathbf{a}\right)\right)\;\;\;\forall\mathbf{b}\in\appa,
\ee
where $\mathbf{a}$ is an arbitrary element in $\appa=\bh$.
Clearly, different choices of $\mathbf{a}$ may lead to the same $T_{\varrho}i_{+}(\mathbf{V}_{\varrho})$.
Then, writing $\mathbf{a}=\mathbf{x} + \imath\mathbf{y}$ with $\mathbf{x},\mathbf{y}\in\appa_{sa}$, we have
\be
T_{\varrho}i_{+}(\mathbf{V}_{\varrho})(\mathbf{b})\,=\,\mathrm{Tr}\left(\left(\left\{\varrho,\,\mathbf{x}\right\} -\imath\,\left[\varrho,\, \mathbf{y}\right]\right)\,\mathbf{b}\right)\;\;\;\forall\mathbf{b}\in\appa,
\ee
with $\{\cdot,\,\cdot\}$ and $[\cdot,\,\cdot]$ the anticommutator and the commutator in $\bh$, respectively.

\subsection{Faithful, finite trace}\label{subsec: faithful, finite trace}

Let $\appa$ be a unital $C^*$-algebra with a faithful, finite trace $\tau$, that is, $\tau$ is a faithful, positive linear functional on $\appa$ such that
\be
\tau(\mathbf{a}\,\mathbf{b})\,=\,\tau(\mathbf{b}\,\mathbf{a})\;\;\;\forall\,\,\mathbf{a},\mathbf{b}\in\appa .
\ee
In particular, if $\appa$ is Abelian, then every faithful, positive linear functional is a faithful, finite trace.
 
We will prove that the orbit $\mathcal{O}_{+}^{\tau}$  of $\gapp$ through $\tau$ by means of the linear action $\alpha$ (see equation \eqref{eqn: linear action of invertible elements on the dual}) is a homogeneous Banach manifold of $\gapp$.
At this purpose, the characterization of an element $\mathbf{a}$ in the Lie algebra of the isotropy group $\gapp_{\tau}$ given in equation \eqref{eqn: isotropy subalgebra for linear action} reads
\be\label{eqn: isotropy algebra of finite trace}
\tau((\mathbf{a} + \mathbf{a}^{\dagger})\,\mathbf{b})\,=\,0\;\;\;\forall \,\,\mathbf{b}\in\appa
\ee
because $\tau$ is a trace, and we see that the skew-adjoint part of $\mathbf{a}$ is completely arbitrary.
Then, we recall that $\tau$ induces an inner product $\langle,\rangle_{\tau}$ on $\appa$ given by\footnote{Note that the faithfulness of $\tau$ is necessary for $\langle,\rangle_{\tau}$ to be an inner product on the whole $\appa$.}
\be
\langle \mathbf{b},\,\mathbf{c}\rangle_{\tau}\,=\,\tau(\mathbf{b}^{\dagger}\,\mathbf{c}).
\ee
Consequently, the completion of $\appa$ with respect to $\langle,\rangle_{\tau}$ is a Hilbert space in which $\appa$ is a dense subspace and thus the validity of equation \eqref{eqn: isotropy algebra of finite trace} implies 
\be
\mathbf{a}+\mathbf{a}^{\dagger}\,=\,\mathbf{0} .
\ee
This means that the Lie algebra $\mathfrak{g}_{\tau}$ of $\gapp_{\tau}$ coincides with the space of skew-adjoint elements in $\mathfrak{g}\cong\appa$, and this subspace is a closed and complemented subspace of $\mathfrak{g}\cong\appa$ whose complement is the space of self-adjoint elements in $\mathfrak{g}\cong\appa$.
From this we conclude that the orbit $\mathcal{O}_{+}^{\tau}$ of $\gapp$ through $\tau$ by means of the linear action $\alpha$ is a homogeneous Banach manifold of $\gapp$.

It is immediate to check that there is a bijection between $\mathcal{O}_{+}^{\tau}$ and the set of positive, invertible elements in $\appa$, that is, elements of the form $\gr\,\gr^{\dagger}$ with $\gr\in\gapp$.
If $\appa$ is finite-dimensional, then $\mathcal{O}_{\tau}$ coincides with the whole space of faithful, positive linear functionals.

\section{State-preserving action of $\gapp$}\label{sec: state-preserving action}

In this section, we will see how to ``deform'' the action of $\gapp$ in such a way that it preserves the space of states $\stsp$.
Indeed, we recall that $\stsp$ is a subset of the space of positive linear functionals $\appa^{*}_{sa}$ characterized by the condition
\be
\rho(\mathbb{I})\,=\,1
\ee
for every $\rho$ in $\stsp$, and thus, if $\rho$ is in $\stsp$, we have
\be
(\alpha(\gr,\,\rho))(\mathbb{I})\,=\,\rho(\gr^{\dagger}\,\gr)
\ee
which is in general different from 1.
To overcome this difficulty, we have to deform the action $\alpha$.
The result is a map $\Phi$ which is not defined for all elements in $\appa^{*}_{sa}$ as it is the case for $\alpha$, but only on the cone of positive linear functionals.
Furthermore, this map $\Phi$ becomes a left action of $\gapp$ only if we restrict it to act on the space of states $\stsp$, and, since $\stsp$ does not posses the structure of Banach manifold as a whole, we can not speak of a smooth action of $\gapp$ on $\stsp$.
However, proceeding  in analogy with what has been done for the case of positive linear functionals, we will see that  the orbits of the action $\Phi$ on $\stsp$ may still be endowed with the structure of smooth homogeneous Banach manifolds of $\gapp$ depending on the behaviour of the isotropy subgroup.

Following what is done in \cite[sec. 6]{grabowski_kus_marmo-symmetries_group_actions_and_entanglement} and \cite[sec. 2]{ciaglia_dicosmo_ibort_laudato_marmo-dynamical_vector_fields_on_the_manifold_of_quantum_states} for the finite-dimensional case $\appa=\mathcal{B}(\hh)$ with $\hh$ being a finite-dimensional, complex Hilbert space,  it is possible to define a map 
\be
\Phi\colon \gapp\,\times\,\stsp\,\longrightarrow\stsp
\ee
setting $\Phi(\gr,\,\rho)\equiv\,\Phi_{\gr}(\rho)$ where $\Phi_{\gr}(\rho)$ acts on $\mathbf{a}\in\appa$ as follows:
\be\label{eqn: non convex action of the group of invertibles on the space of states}
\begin{split}
\left(\Phi_{\gr}(\rho)\right)(\mathbf{a})\,:=\,\frac{\rho(\gr^{\dagger}\,\mathbf{a}\,\gr)}{\rho(\gr^{\dagger}\,\gr)}\,.
\end{split}
\ee
Clearly, this map is well-defined only if the  the term $\rho(\gr^{\dagger}\,\gr)$ in the denominator does not vainish.
This is the content of the following proposition.

\begin{proposition}\label{prop: expectation value of every state on positive invertible elements is positive}
Let $\rho$ be a state on the unital $C^*$-algebra $\appa$, then
\be
\rho(\gr^{\dagger}\,\gr)>0 
\ee
for every element $\gr$ in the group $\gapp$ of invertible elements in $\appa$.

\proof
Let $(\hh_{\rho},\,\pi_{\rho},\,|\psi_{\rho}\rangle)$ be the data of the GNS construction associated with $\rho$ so that
\be
\rho(\gr^{\dagger}\, \gr) \, = \, \langle \psi_{\rho} |\pi_{\rho}(\gr^{\dagger}\,\gr)|\psi_{\rho}\rangle\,.
\ee
The polar decomposition of $\pi_\rho (\gr) $ allows us to write
\be
\pi_\rho (\gr) \,=\, U P
\ee
where $U$ is a unitary operator and $P = \sqrt{ \pi_\rho(\gr)^\dagger \,\pi_\rho (\gr) }=\sqrt{\pi_\rho (\gr^\dagger\,\gr)}$ is a non-negative Hermitean operator.
Then, and because $\gr$ is invertible, we have that $\sqrt{\pi_{\rho}(\gr^\dagger \,\gr)}$ is invertible, which implies that  $\pi_{\rho}(\gr^{\dagger}\,\gr) > 0$ and thus 
\be
\langle \psi_{\rho} |\pi_{\rho}(\gr^{\dagger}\,\gr)|\psi_{\rho}\rangle > 0 .
\ee 
\qed 

\end{proposition}

Direct inspection shows that $\Phi$ is a left action of $\gapp$ on $\stsp$, and that the restriction of the map $\Phi$ to $\uapp \times \stsp$ will define the standard  action of the unitary group $\uapp$ on the space of states $\stsp$: $\left(\Phi_{\gr}(\rho)\right)(\mathbf{a})\,:=\, \rho(\gr^{\dagger}\,\mathbf{a}\,\gr)$.
Furthermore, given a convex combination $\lambda_{1}\,\rho_{1} + \lambda_{2}\,\rho_{2}$ of states on $\appa$, we may have
\be
\Phi_{\gr}(\lambda_{1}\,\rho_{1} + \lambda_{2}\,\rho_{2})\,\neq\,\lambda_{1}\,\Phi_{\gr}(\rho_{1}) + \lambda_{2}\,\Phi_{\gr}(\rho_{2})\,,
\ee
which means that the left action $\Phi$ of $\gapp$ does not preserve the convex structure of $\stsp$ (while that of $\uapp$ does).

In the  work \cite[p. 214]{hellwig_kraus-pure_operations_and_measurements}, the authors take inspiration from the seminal paper \cite[p. 850]{haag_kastler-an_algebraic_approach_to_quantum_field_theory} on the algebraic formulation of quantum field theories to introduce a prototype of the map $\Phi_{\gr}$ in the context of  state transformations and measurements in quantum theories.
However, the map they consider depends on the particular state on which it is applied because they consider the whole algebra $\appa$ instead of the set $\gapp$ of invertible elements thus introducing elements for which the denominator may vanish on the given state (see \cite[sec. 2]{ciaglia_dicosmo_ibort_laudato_marmo-dynamical_vector_fields_on_the_manifold_of_quantum_states} for a finite-dimensional example).

\begin{remark}
The rest of this section is devoted to the study of the action $\Phi$ in complete analogy with what has been done for the action $\alpha$ in section \ref{sec: action}, and we will obtain similar results adopting conceptually similar proofs.
For the sake of completeness, we decided to give a detailed account of all the proofs.
Furthermore, we want to stress a substantial difference between the action $\alpha$ and the action $\Phi$, namely, $\alpha$ is a smooth action on a smooth Banach manifold, while $\Phi$ is just an action on a subset of a Banach manifold.
This means that some of the machinery related with smooth actions (e.g., the notion of fundamental vector field) make no sense in relation with $\Phi$.
\end{remark}

We denote by $\mathcal{O}$ an  orbit of $\gapp$ in $\stsp$ by means of $\Phi$.
Some preliminary characterizations of $\mathcal{O}$ are proved in the following.

\begin{proposition}\label{prop: useful properties of nonconvex action}
Let $\Phi$ be the action of $\gapp$ on $\stsp$ given by equation \eqref{eqn: non convex action of the group of invertibles on the space of states}, then:

\begin{itemize}
\item if $\appa$ is a $W^{*}$-algebra and $\rho$ is a normal state, then $\Phi(\gr,\,\rho)\in\stsp$ is also normal;
\item the state $\Phi(\gr,\,\rho)\in\stsp$ is a faithful state for every $\gr\in\gapp$ if and only if $\rho\in\stsp$ is faithful;
\item the state $\Phi(\gr,\,\rho)\in\stsp$ is a pure state for every $\gr\in\gapp$ if and only if $\rho\in\stsp$ is pure, and if $\appa$ is Abelian, then $\gapp$ acts trivially on pure states by means of $\Phi$;
\item if $\rho$ is a tracial state\footnote{A state $\rho\in\stsp$ is called \grit{tracial} if $\rho(\mathbf{ab})=\rho(\mathbf{ba})$ for all $\mathbf{a},\mathbf{b}\in\appa$. }, then the orbit $\mathcal{O}$ containing $\rho$  is convex; in particular, every orbit $\mathcal{O}$ is convex for all $\rho\in\stsp$ when $\appa$ is Abelian.
\end{itemize}

\proof

Concerning the first point, a normal state $\rho$ is an element of $\stsp$ which is also continuous with respect to the weak* topology on $\appa$ generated by its topological predual $\appa_{*}$.
Recall that, for every $\mathbf{b}\in\appa$, the maps
\be
\begin{split}
l_{\mathbf{b}}\,&\colon\appa\rightarrow\appa,\;\;\;\; l_{\mathbf{b}}(\mathbf{a}):=\mathbf{ba} \\
r_{\mathbf{b}}\,&\colon\appa\rightarrow\appa,\;\;\;\;  r_{\mathbf{b}}(\mathbf{a}):=\mathbf{ab}
\end{split}
\ee
are continuous with respect to the weak* topology on $\appa$ generated by its topological predual $\appa_{*}$.
Furthermore, given any normal state $\rho$ and any invertible element $\gr\in\gapp$, we may define the positive real number 
\be
c_{\rho\gr}:=\rho(\gr^{\dagger}\,\gr).
\ee
Now, let $\alpha\colon \mathbb{R}^{+}\times\mathbb{C}\rightarrow\mathbb{C}$ be the continuous\footnote{The topology on $\mathbb{R}^{+}$ is the Lie group topology, the topology on $\mathbb{C}$ is the norm topology associated with the norm $|z|:=\sqrt{z\overline{z}}$, and the topology on $\mathbb{R}^{+}\times\mathbb{C}$ is the product topology associated with the previous two topologies.} left action of the multiplicative group $\mathbb{R}^{+}$ of positive real numbers on $\mathbb{C}$ given by 
\be
\alpha(c,\,z)\,:=\,cz.
\ee
It is immediate to check that the normalized positive linear functional $\Phi(\gr,\,\rho)\colon\appa\rightarrow\mathbb{C}$ may be written as
\be
\Phi(\gr,\,\rho)\,=\,\alpha_{c_{\rho\gr}^{-1}}\,\circ\,\rho\,\circ\,l_{\gr^{\dagger}}\,\circ\,r_{\gr}\,,
\ee
where $\alpha_{c}(z)=\alpha(c,\,z)$, and thus $\Phi(\gr,\,\rho)$ is weak* continuous. 

The second point follows by direct inspection.

Concerning the third point, let $(\hh_{\rho},\,\pi_{\rho},\,|\psi_{\rho}\rangle)$ be the GNS data associated with a pure state $\rho$.
Then, it is a matter of direct computation to show that $(\hh_{\gamma},\,\pi_{\gamma},\,|\psi_{\gamma}\rangle)$, where $\hh_{\gamma}=\hh_{\rho}$, $\pi_{\gamma}=\pi_{\rho}$ and
\be
|\psi_{\gamma}\rangle\,=\,\frac{\pi_{\rho}(\gr)|\psi_{\rho}\rangle}{\sqrt{\langle\psi_{\rho}|\pi_{\rho}(\gr^{\dagger}\gr)|\psi_{\rho}\rangle}}\,,
\ee
is the data of the GNS construction associated with $\gamma=\Phi(\gr,\,\rho)$.
Since $\rho$ is pure, we have that  $\pi_{\rho} = \pi_{\gamma}$ is irreducible which means that $\gamma$ is pure, and we conclude that the orbit $\mathcal{O}$ containing the pure state $\rho$ is made up only of pure states.
A direct consequence is that the the action of $\gapp$ on the space of pure states of a commutative, unital $C^*$-algebra is trivial in the sense that every pure state is a fixed point of the action.
Indeed, recalling that the GNS representation associated with a pure state $\rho$ is irreducible, then \cite[p. 102]{blackadar-operator_algebras_theory_of_c*-algebras_and_von_neumann_algebras} implies that the GNS Hilbert space $\hh_{\rho}$ is one-dimensional since $\appa$ is commutative.
Consequently, the GNS representation $\pi_{\rho}$ sends every element in the identity operator on $\hh_{\rho}$ and we conclude that the orbit $\mathcal{O}$ containing $\rho$ is just the singleton $\{\rho\}$.
As we will see, this is in sharp contrast with what happens in the non-commutative case.

Regarding the last point, we start taking $\lambda\in[0,1]$, a tracial state $\rho$, two elements $\gr_{1},\gr_{2}\in\gapp$, and writing
\be
\rho_{12}^{\lambda}\,:=\,\lambda\,\Phi(\gr_{1},\rho) + (1-\lambda)\Phi(\gr_{2},\rho).
\ee
Then, for every $\mathbf{a}\in\appa$, we have
\be
\begin{split}
\rho_{12}^{\lambda}(\mathbf{a})& \,=\,\lambda\frac{\rho(\gr_{1}^{\dagger}\,\mathbf{a}\,\gr_{1})}{\rho(\gr_{1}^{\dagger}\,\gr_{1})} + (1-\lambda)\frac{\rho(\gr_{2}^{\dagger}\,\mathbf{a}\,\gr_{2})}{\rho(\gr_{2}^{\dagger}\,\gr_{2})}\,=\, \\
&\,=\, \rho\left(\lambda\frac{\gr_{1}\,\gr_{1}^{\dagger}}{\rho(\gr_{1}\,\gr_{1}^{\dagger})}\,\mathbf{a}\right) + \rho\left((1-\lambda)\frac{\gr_{2}\,\gr_{2}^{\dagger}}{\rho(\gr_{2}\,\gr_{2}^{\dagger})}\,\mathbf{a}\right)\,=\, \\
&\,=\, \rho\left(\left(\lambda\frac{\gr_{1}\,\gr_{1}^{\dagger}}{\rho(\gr_{1}\,\gr_{1}^{\dagger})} + ((1-\lambda)\frac{\gr_{2}\,\gr_{2}^{\dagger}}{\rho(\gr_{2}\,\gr_{2}^{\dagger})}\right)\,\mathbf{a}\right)\,=\, \\
&\,=\,\rho\left(\mathbf{P}_{12}^{\lambda\rho}\,\mathbf{a}\right),
\end{split}
\ee
where we have set
\be
\mathbf{P}_{12}^{\lambda\rho}\,:=\,\lambda \mathbf{P}_{1}^{\rho} + (1-\lambda) \mathbf{P}_{1}^{\rho}
\ee
with
\be
\mathbf{P}_{1}^{\rho}\,:=\,\frac{\gr_{1}\,\gr_{1}^{\dagger}}{\rho(\gr_{1}\,\gr_{1}^{\dagger})}  \mbox{ and } \;\mathbf{P}_{2}^{\rho}\,:=\,\frac{\gr_{2}\,\gr_{2}^{\dagger}}{\rho(\gr_{2}\,\gr_{2}^{\dagger})}\,.
\ee
The elements $\mathbf{P}_{1}^{\rho}\,:=\,\frac{\gr_{1}\,\gr_{1}^{\dagger}}{\rho(\gr_{1}\,\gr_{1}^{\dagger})}$ and $\mathbf{P}_{2}^{\rho}$ are both positive, invertible elements in $\appa$, and we have that the set
\be
\gapp_{+}\,=\,\gapp\,\cap\,\appa_{+}
\ee
of positive, invertible elements (strictly positive elements) in $\appa$ is an open cone  (see \cite[p. 11]{takesaki-theory_of_operator_algebra_I}), so that $\mathbf{P}_{12}^{\lambda\rho}$ is still a positive, invertible element.
Being a positive element, $\mathbf{P}_{12}^{\lambda\rho}$ admits a (self-adjoint) square root, say $\mathbf{p}_{12}^{\lambda\rho}$, and this square-root element is also invertible, i.e., $\mathbf{p}_{12}^{\lambda\rho}\in\gapp$.
Consequently, noting that 
\be
\rho\left(\mathbf{P}_{12}^{\rho}\right)\,=\,1,
\ee
we obtain
\be
\rho_{12}^{\lambda}(\mathbf{a})\,=\,\frac{\rho\left((\mathbf{p}_{12}^{\lambda\rho})^{\dagger}\,\mathbf{a}\,\mathbf{p}_{12}^{\lambda\rho}\right)}{\rho\left((\mathbf{p}_{12}^{\lambda\rho})^{\dagger}\,\mathbf{p}_{12}^{\lambda\rho}\right))}
\ee
for all $\mathbf{a}\in\appa$.
This is equivalent to
\be
\rho_{12}^{\lambda}\,=\,\Phi(\mathbf{p}_{12}^{\lambda\rho},\rho),
\ee
which means that the orbit $\mathcal{O}$ containing $\rho$ is convex as claimed.
In particular, every orbit $\mathcal{O}$ is convex  when $\appa$ is Abelian  because all states are tracial.
\qed

\end{proposition}

Let $\mathcal{O}\subset\stsp$ be an orbit of $\gapp$ by means of $\Phi$.
Considering $\rho\in\mathcal{O}$ and the coset space $\gapp/\gapp_{\rho}$, where $\gapp_{\rho}$ is the isotropy subgroup
\be
\gapp_{\rho}\,=\,\left\{\gr\in\gapp\,\colon\:\:\Phi(\gr,\,\rho)\,=\, \rho \right\} ,
\ee 
of $\rho$ with respect to $\Phi$,  the map $i_{\rho}^{\Phi}\colon \gapp/\gapp_{\rho}\rightarrow \mathcal{O}$ given by
\be
[\gr]\mapsto i_{\rho}^{\Phi}([\gr])=\Phi(\gr,\rho)
\ee
provides a set-theoretical bijection between the coset space $\gapp/\gapp_{\rho}$   and the orbit $\mathcal{O}$ for every $\rho\in\stsp$.
According to the results recalled in appendix \ref{app: banach-lie groups and homogeneous spaces}, this means that  we may dress the orbit $\mathcal{O}$ with the structure of homogeneous Banach manifold of $\gapp$ whenever the isotropy subgroup $\gapp_{\rho}$ is a Banach-Lie subgroup of $\gapp$.
Specifically, it is the quotient space $\gapp/\gapp_{\rho}$ that is endowed with the structure of homogeneous Banach manifold, and this structure may be ``transported'' to $\mathcal{O}$  in view of the bijection $ i_{\rho}^{\Phi}$ between $\gapp/\gapp_{\rho}$ and $\mathcal{O}$.

As it happens for the action $\alpha$ defined in section \ref{sec: action}, in general, the fact that $\gapp_{\rho}$ is a Banach-Lie subgroup of $\gapp$ depends on both $\rho$ and $\appa$.
However,  $\gapp_{\rho}$ is always an algebraic subgroup of $\gapp$ for every $\rho\in\stsp$ and every unital $C^*$-algebra $\appa$ (see the discussion above proposition \ref{prop: isotropy subgroups of linear action on positive linear functionals are algebraic subgroups} for the definition and the properties of algebraic subgroups of a Banach-Lie group).

\begin{proposition}\label{prop: isotropy subgroups are algebraic subgroups}
The isotropy subgroup  $\gapp_{\rho}$ of $\rho\in\stsp$ is an algebraic subgroup of $\gapp$ of order $2$ for every $\rho\in\stsp$.

\proof
The proof is essentially the same of proposition \ref{prop: isotropy subgroups of linear action on positive linear functionals are algebraic subgroups} with only a slight modification of the family of polynomials considered.
Define the family $Q_{\rho}=\{p_{\rho,\mathbf{c}}\}_{\mathbf{c}\in\appa}$ of complex-valued polynomials of order $2$ as follows\footnote{Note that the dependence of $p_{\rho,\mathbf{c}}$ on the second variable is trivial, and this explains why $\mathbf{b}$ does not appear on the rhs.}:
\be
p_{\rho,\mathbf{c}}(\mathbf{a}\,,\mathbf{b})\,:=\,\rho(\mathbf{a}^{\dagger}\,\mathbf{a})\,\rho\left(\mathbf{c}\right) \,-\, \rho\left(\mathbf{a}^{\dagger}\,\mathbf{c}\mathbf{a}\right)\,.
\ee 
The continuity of every $p_{\rho,\mathbf{c}}$ follows easily from the fact that $\rho$ is a norm-continuous linear functional on $\appa$.
A moment of reflection shows that
\be
\gapp_{\rho}=\left\{\gr\in\gapp \colon p_{\rho,\mathbf{c}}(\gr\,,\gr^{-1})=0\;\;\forall p_{\rho,\mathbf{c}}\in Q_{\rho}\right\}\,,
\ee
and thus $\gapp_{\rho}$ is an algebraic subgroup of $\gapp$ of order $2$ for all $\rho\in\stsp$. 
\qed
\end{proposition} 

Being an algebraic subgroup of $\gapp$, the isotropy subgroup $\gapp_{\rho}$ is  a closed subgroup of $\gapp$ which is also a real Banach-Lie group in the relativised norm topology, and its Lie algebra $\mathfrak{g}_{\rho}\subset\mathfrak{g}=\appa$ is given by the closed subalgebra (see  \cite[p. 667]{harris_kaup-linear_algebraic_groups_in_infinite_dimensions}, and \cite[p. 118]{upmeier-symmetric_banach_manifolds_and_jordan_calgebras})
\be
\mathfrak{g}_{\rho}=\left\{\mathbf{a}\in\mathfrak{g}\equiv\appa\,\colon \exp(t\mathbf{a})\in\gapp_{\rho}\;\forall t\in\mathbb{R}\right\}\,.
\ee
According to proposition \ref{prop: characterization of Banach-Lie subgroups}, the isotropy subgroup $\gapp_{\rho}$ of $\gapp$ is a Banach-Lie  subgroup of $\gapp$ if and only if the Lie algebra $\mathfrak{g}_{\rho}$ of $\gapp_{\rho}$ is a split subspace of $\mathfrak{g}=\appa$ and $\exp(V)$ is a neighbourhood of the identity element in $\mathcal{G}_{\rho}$ for every neighbourhood $V$ of $\mathbf{0}\in\mathfrak{g}_{\rho}$ (see \cite[p. 129]{upmeier-symmetric_banach_manifolds_and_jordan_calgebras} for an explicit proof).
The fact that $\exp(V)$ is a neighbourhood of the identity element in $\gapp_{\rho}$ for every neighbourhood $V$ of $\mathbf{0}\in\mathfrak{g}_{\rho}$ follows from the fact that $\gapp_{\rho}$ is an algebraic subgroup of $\gapp$ (see \cite[p. 667]{harris_kaup-linear_algebraic_groups_in_infinite_dimensions}).

Next, we may characterize $\mathfrak{g}_{\rho}$ as we did in section \ref{sec: action} by considering $\mathbf{a}\in \mathfrak{g}=\appa$, the smooth curve in $\gapp$ given by
\be
\gr_{t}\,=\,\exp(t\mathbf{a}) 
\ee
for all $t\in\mathbb{R}$,  the curve $\rho_{t}$ in $\stsp$ given by
\be
\rho_{t}(\mathbf{b})\,=\,(\Phi(\gr_{t},\,\rho))(\mathbf{b})\,=\,\rho\left(\gr_{t}^{\dagger}\,\mathbf{b}\,\gr_{t}\right) 
\ee
for all $t\in\mathbb{R}$ and for all $\mathbf{b}\in\appa$, and computing
\be\label{eqn: isotropy subalgebra of Phi}
\begin{split}
\frac{\mathrm{d}}{\mathrm{d}t}\,\left(\rho_{t}(\mathbf{b})\right)_{t=0}&\,=\,\lim_{t\ra 0}\, \frac{\rho_{t}(\mathbf{b}) -\rho(\mathbf{b})}{t}\,=\,\\
&\,=\, \lim_{t\ra 0}\, \frac{1}{t}\,\left(\frac{\rho(\gr_{t}^{\dagger}\,\mathbf{b}\,\gr_{t})}{\rho(\gr_{t}^{\dagger}\,\gr_{t})} - \rho(\mathbf{b})\right)\,=\,\\
&\,=\,\lim_{t\ra 0}\, \frac{1}{t\,\rho(\gr_{t}^{\dagger}\,\gr_{t})}\,\left( \rho\left(\gr_{t}^{\dagger}\,(\mathbf{b}- \rho(\mathbf{b})\,\mathbb{I})\,\gr_{t}\right)\right) \,=\,\\
&\,=\,\lim_{t\ra 0}\, \frac{1}{t}\,\left( \rho\left(\gr_{t}^{\dagger}\,(\mathbf{b}- \rho(\mathbf{b})\,\mathbb{I})\,\gr_{t}\right)\right)\,\left(\lim_{t\ra 0}\, \frac{1}{\rho(\gr_{t}^{\dagger}\,\gr_{t})}\right)\,=\, \\
&\,=\,\lim_{t\ra 0}\, \frac{1}{t}\,\left( \rho\left(\gr_{t}^{\dagger}\,(\mathbf{b}- \rho(\mathbf{b})\,\mathbb{I})\,\gr_{t}\right)\right)\,=\,\\
&\,=\,\lim_{t\ra 0}\, \frac{1}{t}\,\sum_{j,k=0}^{+\infty}\,\left( \rho\left(\frac{(t\mathbf{a}^{\dagger})^{k}}{k!}\,(\mathbf{b}- \rho(\mathbf{b})\,\mathbb{I})\,\frac{(t\mathbf{a})^{j}}{j!}\right)\right) \,=\,\\
&\,=\,\rho\left(\mathbf{a}^{\dagger}\,(\mathbf{b}- \rho(\mathbf{b})\,\mathbb{I})\right)\, + \rho\left((\mathbf{b}- \rho(\mathbf{b})\,\mathbb{I})\,\mathbf{a}\right)\,=\,\\
&\,=\,\rho\left(\mathbf{a}^{\dagger}\,\mathbf{b}\,+\,\mathbf{b}\,\mathbf{a}\right)\,-\,\rho(\mathbf{b})\,\rho\left(\mathbf{a}^{\dagger}\,+\,\mathbf{a}\right)
\end{split}
\ee
for every $\mathbf{b}\in\appa$, from which it follows that $\mathbf{a}$ is in the Lie algebra $\mathfrak{g}_{\rho}$ of the isotropy group $\gapp_{\rho}$ if and only if
\be\label{eqn: isotropy subalgebra of Phi 2}
\rho\left(\mathbf{a}^{\dagger}\,\mathbf{b}\,+\,\mathbf{b}\,\mathbf{a}\right)\,-\,\rho(\mathbf{b})\,\rho\left(\mathbf{a}^{\dagger}\,+\,\mathbf{a}\right)\,=\,0
\ee
for every $\mathbf{b}\in\appa$.
Incidentally, note that the last term in equation \eqref{eqn: isotropy subalgebra of Phi} gives the covariance between $\mathbf{b}$ and $\mathbf{a}$ evaluated at the state $\rho$ whenever $\mathbf{a}$ is self-adjoint.
Something related has also been pointed out in \cite[eqn. 34]{chruscinski_ciaglia_ibort_marmo_ventriglia-stratified_manifold_of_quantum_states}, and we postpone to a future work a more thorough analysis of the connection between the action of $\gapp$ on $\stsp$ and the existence of contravariant tensor fields associated with the covariance between observables (in the $C^*$-algebraic sense).

When $\mathrm{dim}(\appa)=N<\infty$, the Lie algbera $\mathfrak{g}_{\rho}$ is a split subspace for ever $\rho\in\stsp$, and thus every orbit of $\gapp$ in $\stsp$ by means of $\Phi$ is a homogeneous Banach manifold of $\gapp$.
Clearly, when $\appa$ is infinite-dimensional, this is no-longer true, and a case by case analysis is required.
For instance, in subsection \ref{sec: density operators}, we will show that $\mathfrak{g}_{\rho}$ is a split subspace of $\mathfrak{g}=\appa$ when $\appa$ is the algebra $\bh$ of bounded linear operators on a complex, separable Hilbert space $\hh$, and $\rho$ is any normal state on $\bh$ (positive, trace-class linear operator on $\hh$ with unit trace).
This means that all the orbits of $\gapp=\mathcal{GL}(\hh)$ passing through normal states are homogeneous Banach manifolds of $\gapp$, and we will classify these orbits into four different types.

Actually, the results of subsection \ref{sec: density operators} naturally follows from the results of subsection \ref{sebsec: positive trace-class operators} because, as we will now show, there is an intimate connection between the action $\alpha$ of $\gapp$ on $\rho$ when the latter is thought of as an element of $\appa^{*}_{sa}$, and the action $\Phi$ of $\gapp$ on $\rho$ when the latter is thought of as an element of $\stsp$.
Indeed, from equations \eqref{eqn: linear action of invertible elements on the dual} and \eqref{eqn: non convex action of the group of invertibles on the space of states}, we easily obtain that if $\gr$ is in the isotropy group $\gapp_{\rho}^{\alpha}$ of $\rho$ with respect to the action $\alpha$, then $\gr$ is also in the isotropy group $\gapp_{\rho}$ of $\rho$ with respect to $\Phi$, while the converse is not necessarily true.
Furthermore, if $\gr$ is in $\gapp^{\alpha}_{\rho}$, then $\mathrm{e}^{\gamma}\gr$ is in $\gapp_{\rho}$ for every $\gamma\in\mathbb{R}$, and it turns out that this is the most general expression for an element in $\gapp_{\rho}$.
This is made precise in the following proposition where we show that the Lie algebra $\mathfrak{g}_{\rho}$ of $\gapp_{\rho}$  is just the direct sum of the Lie algebra $\mathfrak{g}_{\rho}^{\alpha}$ with the one-dimensional subspace determined by the linear combinations of multiples of the identity with real coefficients.

\begin{proposition}\label{prop: isotropy subalgebra of Phi is the isotropy subalgebra of alpha plus the identity with real coefficients}
The Lie algebra $\mathfrak{g}_{\rho}$ of the isotropy group $\gapp_{\rho}$ of $\rho$ with respect to $\Phi$ may be written as
\be
\mathfrak{g}_{\rho}\,=\,\mathfrak{g}_{\rho}^{\alpha}\,\oplus\,\mathrm{span}_{\mathbb{R}}\{\mathbb{I}\}
\ee
where $\mathfrak{g}_{\rho}^{\alpha}$ is the Lie algebra of the isotropy group $\gapp_{\rho}^{\alpha}$ of $\rho$ with respect to the action $\alpha$ introduced in section \ref{sec: action}, and $\mathrm{span}_{\mathbb{R}}\{\mathbb{I}\}$ is the real, linear subspace spanned by the identity operator in $\gapp=\appa$ with real coefficients.

\proof
It is a matter of direct inspection to see that if $\mathbf{a}$ is in $\mathfrak{g}_{\rho}^{\alpha}$ (see equation \eqref{eqn: isotropy subalgebra for linear action}), then $\mathbf{a} + \gamma\mathbb{I}$ is in $\mathfrak{g}_{\rho}$ for every $\gamma\in\mathbb{R}$.

On the other hand, since $\mathrm{span}_{\mathbb{R}}\{\mathbb{I}\}$ is one-dimensional, it is complemented in $\mathfrak{g}_{\rho}$, and we may characterize its complement as follows.
First, we take the continuous, real linear functional $F$ on $\mathrm{span}_{\mathbb{R}}\{\mathbb{I}\}$ given by
\be
F(\gamma\,\mathbb{I})\,:=\,\gamma,
\ee
and extend it to the whole $\mathfrak{g}_{\rho}$.
The extension of $F$ is highly non-unique, and we may take it to be the functional $F_{\rho}$ given by 
\be
F_{\rho}(\mathbf{a})\,:=\,\frac{1}{2}\,(\rho + \rho^{\dagger})(\mathbf{a})\,=\,\frac{1}{2}\,\rho(\mathbf{a}^{\dagger} + \mathbf{a}) \;\;\forall\,\,\mathbf{a}\in\mathfrak{g}_{\rho}.
\ee
Indeed, $F_{\rho}$ is a real, continuous linear functional on $\mathfrak{g}_{\rho}$ because $(\rho + \rho^{\dagger})$ is a continuous linear functional on the real Banach-Lie algebra $\mathfrak{g}=\appa$ of which $\mathfrak{g}_{\rho}$ is a closed, real subalgebra, and clearly $F_{\rho}(\gamma\mathbb{I})\,=\,F(\gamma\mathbb{I})$ because $\rho$ is a state.
Then, we have a bounded projection $P$ from $\mathfrak{g}_{\rho}$ to $\mathrm{span}_{\mathbb{R}}\{\mathbb{I}\}$ given by
\be
P(\mathbf{a})\,=\,F_{\rho}(\mathbf{a})\,\mathbb{I}\;\;\forall\,\,\mathbf{a}\in\mathfrak{g}_{\rho},
\ee
and we may write
\be
\mathbf{a}\,=\, P(\mathbf{a}) - \left(\mathrm{Id}_{\mathfrak{g}_{\rho}} - P\right)(\mathbf{a})\;\;\forall\,\,\mathbf{a}\in\mathfrak{g}_{\rho}.
\ee
This allows us to define the complement of $\mathrm{span}_{\mathbb{R}}\{\mathbb{I}\}$ in $\mathfrak{g}_{\rho}$ as the closed linear subspace $\mathfrak{c}_{\rho}$ given by the image of $\left(\mathrm{Id}_{\mathfrak{g}_{\rho}} - P\right)$.
Equivalently, an element $\mathbf{b}\in\mathfrak{c}_{\rho}$ may be written as
\be
\mathbf{b}\,=\,\left(\mathrm{Id}_{\mathfrak{g}_{\rho}} - P\right)(\mathbf{a})
\ee
with $\mathbf{a}\in\mathfrak{g}_{\rho}$.
All that is left to do is to show that $\mathbf{b}\in\mathfrak{c}_{\rho}$ is actually in $\mathfrak{g}_{\rho}^{\alpha}$.
At this purpose, recalling equation \eqref{eqn: isotropy subalgebra for linear action}, we have
\be
\begin{split}
\rho(\mathbf{b}^{\dagger}\,\mathbf{c}+\mathbf{c}\,\mathbf{b})&\,=\,\rho\left((\mathbf{a} - \frac{1}{2}\,\rho(\mathbf{a}^{\dagger} + \mathbf{a})\,\mathbb{I})^{\dagger}\,\mathbf{c}+\mathbf{c}\,(\mathbf{a} - \frac{1}{2}\,\rho(\mathbf{a}^{\dagger} + \mathbf{a})\,\mathbb{I})\right)\,=\,\\
&\,=\,\rho(\mathbf{a}^{\dagger}\,\mathbf{c} + \mathbf{c}\,\mathbf{a})- \rho(\mathbf{c})\,\rho(\mathbf{a}^{\dagger} + \mathbf{a})\,=\,0
\end{split}
\ee
because $\mathbf{a}$ is in $\mathfrak{g}_{\rho}$ (see equation \eqref{eqn: isotropy subalgebra of Phi 2}).

\qed
 
\end{proposition}

From proposition \ref{prop: isotropy subalgebra of Phi is the isotropy subalgebra of alpha plus the identity with real coefficients} it follows that $\gapp_{\rho}$ is a Banach-Lie subgroup of $\gapp$ if and only if $\gapp_{\rho}^{\alpha}$ is a Banach-Lie subgroup of $\gapp$.
Consequently, the orbit of $\gapp$ through $\rho$ by means of $\alpha$ is a homogeneous Banach manifold of $\gapp$ if and only if the orbit of $\gapp$ through $\rho$ by means of $\Phi$ is a homogeneous Banach manifold of $\gapp$.

\begin{proposition}\label{prop: isotropy group of Phi is Banach-Lie subgroup iff isotropy group of alpha is}
The Lie algebra $\mathfrak{g}_{\rho}$ is a split subspace of $\mathfrak{g}=\appa$ if and only if the Lie algebra $\mathfrak{g}_{\rho}^{\alpha}$ is so.

\proof
According to proposition \ref{prop: isotropy subalgebra of Phi is the isotropy subalgebra of alpha plus the identity with real coefficients}  we may write
\be
\mathfrak{g}_{\rho}\,=\,\mathfrak{g}_{\rho}^{\alpha}\,\oplus\,\mathrm{span}_{\mathbb{R}}\{\mathbb{I}\},
\ee
Consequently, if $\mathfrak{g}_{\rho}$ is complemented in $\mathfrak{g}=\appa$, we have
\be
\mathfrak{g}\,=\,\mathfrak{g}_{\rho}\,\oplus\,\mathfrak{k}_{\rho}
\ee
and thus the closed linear subspace $\mathfrak{k}_{\rho}\,\oplus\,\mathrm{span}_{\mathbb{R}}\{\mathbb{I}\}$ provides a closed complement for $\mathfrak{g}_{\rho}^{\alpha}$.
On the other hand, if $\mathfrak{g}_{\rho}^{\alpha}$ is complemented in $\mathfrak{g}=\appa$, and we may write
\be
\mathfrak{g}\,=\,\mathfrak{g}_{\rho}^{\alpha}\,\oplus\,\mathfrak{k}_{\rho}^{\alpha}.
\ee
Then, recall that $\gamma\mathbb{I}$ with $\gamma\in\mathbb{R}$ is in $\mathfrak{g}_{\rho}^{\alpha}$ if and only if $\gamma=0$ (see equation \eqref{eqn: isotropy subalgebra for linear action}), therefore, the closed one-dimensional subspace $\mathrm{span}_{\mathbb{R}}\{\mathbb{I}\}$ is a closed linear subspace of $\mathfrak{k}_{\rho}^{\alpha}$, and it is complemented in $\mathfrak{k}_{\rho}^{\alpha}$ because it is finite-dimensional.
Denoting by $\mathfrak{c}_{\rho}$ the complement of $\mathrm{span}_{\mathbb{R}}\{\mathbb{I}\}$ in $\mathfrak{k}_{\rho}^{\alpha}$, we have that 
\be
\mathfrak{g}\,=\,\mathfrak{g}_{\rho}^{\alpha}\,\oplus\,\mathrm{span}_{\mathbb{R}}\{\mathbb{I}\}\,\oplus\,\mathfrak{c}_{\rho}^{\alpha}\,=\,\mathfrak{g}_{\rho}\,\oplus\,\mathfrak{c}_{\rho}^{\alpha},
\ee
from which it follows that $\mathfrak{g}_{\rho}$ is complemented in $\mathfrak{g}=\appa$.
\qed
\end{proposition}

Now, \grit{suppose} $\rho$ is such that $\mathfrak{g}_{\rho}$ is a split subspace of $\appa$, that is, the isotropy subgroup $\gapp_{\rho}$ is a Banach-Lie subgroup of $\gapp$.
In this case, the orbit $\mathcal{O}$ containing $\rho$ is endowed with a Banach manifold structure such that the map $\tau_{\rho}^{\Phi}\colon\gapp\rightarrow \mathcal{O}$ given by
\be\label{eqn: submersion for Phi}
\gr\,\mapsto\,\tau_{\rho}^{\Phi}(\gr)\,:=\,\Phi(\gr,\rho)
\ee
is a smooth surjective submersion for every $\rho\in\mathcal{O}$.
Moreover, $\gapp$ acts transitively and smoothly on $\mathcal{O}$, and  the tangent space $T_{\rho}\mathcal{O}$ at $\rho\in\mathcal{O}$ is diffeomorphic to $\mathfrak{g}/\mathfrak{g}_{\rho}$ (see \cite[p. 105]{bourbaki-groupes_et_algebres_de_lie} and  \cite[p. 136]{upmeier-symmetric_banach_manifolds_and_jordan_calgebras}).
Note that this smooth differential structure on  $\mathcal{O}$  is unique up to smooth diffeomorphism.
Now, we will prove a proposition very similar to proposition \ref{prop: canonical immersion of the orbits through positive linear functionals is smooth} in section \ref{sec: action}.

\begin{proposition}\label{prop: canonical immersion of an orbit in the dual is smooth}
Let $\rho$ be such that the isotropy subgroup $\gapp_{\rho}$ is a Banach-Lie subgroup of $\gapp$, let $\mathcal{O}$ be the orbit containing $\rho$ endowed with the smooth structure coming from $\gapp$, and consider the map $l_{\mathbf{a}}\colon\,\mathcal{O}\lra\mathbb{R}$, with $\mathbf{a}$ a self-adjoint element in $\appa$, given by
\be
l_{\mathbf{a}}(\rho)\,:=\,\rho(\mathbf{a}).
\ee
Then:
\begin{enumerate}
\item the canonical immersion map $i\colon\mathcal{O}\lra\appa^{*}_{sa}$ is  smooth;
\item the map  $l_{\mathbf{a}}\colon\,\mathcal{O}\lra\mathbb{R}$ is smooth;
\item the tangent map $T_{\rho}i$ at $\rho\in\mathcal{O}$ is injective for all $\rho$ in the orbit.
\end{enumerate}

\proof
\begin{enumerate}
\item We will exploit proposition \ref{prop: smoothness of maps from homogeneous space is related with  smoothness of maps from group} in appendix \ref{app: banach-lie groups and homogeneous spaces} in order to prove the smoothness of the canonical immersion.
Specifically, we consider the map
\be
\Phi_{\rho}\,\colon\,\gapp\,\lra\,\appa_{sa}^{*},\;\;\;\Phi_{\rho}(\gr)\,:=\,\Phi(\gr,\rho)
\ee
and note that, quite trivially, it holds
\be
\Phi_{\rho}\,=\,i\,\circ\,\tau_{\rho}^{\Phi}.
\ee
Consequently, being $\tau_{\rho}^{\Phi}$  a smooth submersion for every $\rho\in\stsp$,  proposition \ref{prop: smoothness of maps from homogeneous space is related with  smoothness of maps from group}  implies that $i$ is smooth if $\Phi_{\rho}$ is smooth.

At this purpose, given $\mathbf{a},\mathbf{b}\in\appa$ and  $\xi\in\appa^{*}_{sa}$, we define $\xi_{\mathbf{a}\mathbf{b}}\in\appa^{*}_{sa}$ to be 
\be
\xi_{\mathbf{a}\mathbf{b}}(\mathbf{c})\,:=\,\frac{1}{2}\,\left(\xi(\mathbf{a}^{\dagger}\,\mathbf{c}\,\mathbf{b}) + \xi(\mathbf{b}^{\dagger}\,\mathbf{c}\,\mathbf{a})\right)\;\;\forall\,\mathbf{c}\,\in\,\appa_{sa},
\ee
the map $\phi\colon \appa\times \appa^{*}_{sa}\rightarrow \mathbb{R}\times\appa^{*}_{sa}$ given by
\be
\phi(\mathbf{a},\,\xi)\,:=\,(\xi_{\mathbf{a\,a}}(\mathbb{I}),\,\xi_{\mathbf{a\,a}}),
\ee
and the map $P\colon\,(\appa\times \appa^{*}_{sa})\times(\appa\times \appa^{*}_{sa})\times(\appa\times \appa^{*}_{sa})\rightarrow \mathbb{R}\times\appa^{*}_{sa}$ given by
\be
 F(\mathbf{a},\rho\,;\mathbf{b},\sigma\,;\mathbf{c},\tau)\,:=\,\left(\frac{1}{3}\left(\xi_{\mathbf{bc}}(\mathbb{I}) + \zeta_{\mathbf{ca}}(\mathbb{I}) + \vartheta_{\mathbf{ab}}(\mathbb{I})\right),\,\frac{1}{3}\left(\xi_{\mathbf{bc}} + \zeta_{\mathbf{ca}} + \vartheta_{\mathbf{ab}}\right)\right)\,.
\ee
A direct computation shows that $F$ is a bounded multilinear map and that
\be
\phi(\mathbf{a},\,\xi)\,=\,F(\mathbf{a},\xi\,;\mathbf{a},\xi\,;\mathbf{a},\xi)\,,
\ee
which means that $\phi$ is a continuous polynomial map between $\appa\times\appa^{*}_{sa}$ and $\mathbb{R}\times\appa^{*}_{sa}$, hence, it is smooth with respect to the  Banach manifold structures of $\appa\times\appa^{*}_{sa}$ and $\mathbb{R}\times\appa^{*}_{sa}$ (see \cite[p. 63]{chu-jordan_structures_in_geometry_and_analysis}).
Then, we note that $\gapp$ is an open Banach submanifold of $\appa$ (see \cite[p. 96]{upmeier-symmetric_banach_manifolds_and_jordan_calgebras}), and thus the map 
\be
I_{\xi}\,\colon\,\gapp\,\lra\,\appa\,\times\,\appa^{*}_{sa},\;\;\;\;I_{\xi}(\gr)\,:=\,(\gr,\,\xi)
\ee
is a smooth map for every $\xi\in\appa^{*}_{sa}$ so that $\phi\circ\,I_{\xi}$ is a smooth map between $\gapp$ and $\mathbb{R}\,\times\,\appa^{*}_{sa}$ for every $\xi\in\appa^{*}_{sa}$.

In particular, $I_{\rho}$ is smooth for every $\rho\in\stsp$, and its image is in the open submanifold $\mathbb{R}_{0}\,\times\,\appa^{*}_{sa}$ of $\mathbb{R}\,\times\,\appa^{*}_{sa}$.
Therefore, considering the smooth map $\beta\colon\,\mathbb{R}_{0}\times\appa^{*}_{sa}\,\rightarrow\,\appa^{*}_{sa}$ given by
\be
\beta(b,\,\xi)\,:=\,\frac{1}{b}\,\xi\,,
\ee
it follows that $\beta\,\circ\,\phi\circ\,I_{\rho}$ is a smooth map between $\gapp$ and $\appa^{*}_{sa}$ for every $\rho\in\stsp$, and a direct computation shows that
\be
\Phi_{\rho}\,=\,\beta\,\circ\,\phi\circ\,I_{\rho}.
\ee
From this it follows that $\Phi_{\rho}$ is smooth which means that the canonical immersion $i\colon\mathcal{O}\subset\stsp\lra\appa_{sa}^{*}$ is smooth  because of proposition \ref{prop: smoothness of maps from homogeneous space is related with  smoothness of maps from group}.

\item It suffices to note that $l_{\mathbf{a}}$ is the composition of the linear (and thus smooth) map $L_{\mathbf{a}}\colon\appa_{sa}^{*}\lra\mathbb{R}$ given by
\be
L_{\mathbf{a}}(\xi)=\xi(\mathbf{a})
\ee
with the canonical immersion $i$ which is smooth because of what has been proved above.

\item  Now, consider the family $\{l_{\mathbf{a}}\}_{\mathbf{a}\in\appa}$ of smooth functions on the orbit $\mathcal{O}$, and suppose that $\mathbf{V}_{\rho}$ and $\mathbf{W}_{\rho}$ are tangent vectors at $\rho\in\mathcal{O}$  such that
\be\label{eqn: equation for tangent vectors, states}
\langle (\mathrm{d}l_{\mathbf{a}})_{\rho};\,\mathbf{V}_{\rho}\rangle\,=\,\langle (\mathrm{d}l_{\mathbf{a}})_{\rho};\,\mathbf{W}_{\rho}\rangle
\ee
for every $\mathbf{a}\in\appa_{sa}$.
Then, since $l_{\mathbf{a}}=L_{\mathbf{a}}\circ\,i$, we have
\be
\langle (\mathrm{d}l_{\mathbf{a}})_{\rho};\,\mathbf{V}_{\rho}\rangle\,=\,\langle (\mathrm{d}L_{\mathbf{a}})_{i(\rho)};\,T_{\rho}i(\mathbf{V}_{\rho})\rangle
\ee
and
\be
\langle (\mathrm{d}l_{\mathbf{a}})_{\rho};\,\mathbf{W}_{\rho}\rangle\,=\,\langle (\mathrm{d}L_{\mathbf{a}})_{i(\rho)};\,T_{\rho}i(\mathbf{W}_{\rho})\rangle
\ee
Note that the family of linear functions of the type $L_{\mathbf{a}}$ with  $\mathbf{a}\in\appa_{sa}$ (see equation \eqref{eqn: linear functions associated with elements in the predual}) are enough to separate the tangent vectors at $\xi$ for every $\xi\in\appa^{*}_{sa}$ because the tangent space at $\xi\in\appa^{*}_{sa}$ is diffeomorphic with $\appa^{*}_{sa}$ in such a way that 
\be
\langle (\mathrm{d}L_{\mathbf{a}})_{\xi};\mathbf{V}_{\xi}\rangle\,=\,\mathbf{V}_{\xi}(\mathbf{a})\,=\,L_{\mathbf{a}}(\mathbf{V}_{\xi})
\ee
for every $\mathbf{V}_{\xi}\in T_{\xi}\appa^{*}_{sa}\cong\appa^{*}_{sa}$, 
and $\appa_{sa}$ (the predual of $\appa^{*}_{sa}$) separates the points of $\appa^{*}_{sa}$ (see \cite{kaijser-a_note_on_dual_banach_spaces}).
Consequently, since $T_{\rho}i(\mathbf{V}_{\rho})$ and $T_{\rho}i(\mathbf{W}_{\rho})$ are tangent vectors at $i(\rho)\in\appa^{*}_{sa}$ and the functions $L_{\mathbf{a}}$ with $\mathbf{a}\in\appa_{sa}$ are enough to separate them and we conclude that the validity of equation \eqref{eqn: equation for tangent vectors, states} for all $\mathbf{a}\in\appa_{sa}$ is equivalent to 
\be
T_{\rho}i(\mathbf{V}_{\rho})\,=\,T_{\rho}i(\mathbf{W}_{\rho}).
\ee
Then, if $\gr_{t}\,=\,\exp(t\mathbf{a})$ is a one-parameter subgroup in $\gapp$ so that
\be
\rho_{t}\,=\,\Phi(\gr_{t},\,\rho)
\ee
is a smooth curve in $\mathcal{O}$ starting at $\rho$ with associated tangent vector $\mathbf{V}_{\rho}$,   we have
\be
\langle (\mathrm{d}L_{\mathbf{b}})_{i(\rho)};\,T_{\rho}i(\mathbf{V}_{\rho})\rangle\,=\,\frac{\mathrm{d}}{\mathrm{d}t}\,\left(L_{\mathbf{b}}\,\circ\,i(\rho_{t})\right)_{t=0}
\ee
which we may compute analogously to equation \eqref{eqn: isotropy subalgebra of Phi} to obtain
\be\label{eqn: tangent vector at a point in the orbit of Phi}
\begin{split}
\frac{\mathrm{d}}{\mathrm{d}t}\,\left(L_{\mathbf{b}}\,\circ\,i(\rho_{t})\right)_{t=0}&\,=\,\rho\left(\mathbf{a}^{\dagger}\,\mathbf{b}\,+\,\mathbf{b}\,\mathbf{a}\right)\,-\,\rho(\mathbf{b})\,\rho\left(\mathbf{a}^{\dagger}\,+\,\mathbf{a}\right)
\end{split}
\ee
Comparing equation \eqref{eqn: tangent vector at a point in the orbit of Phi} with equation \eqref{eqn: isotropy subalgebra of Phi} we conclude that $\mathbf{V}_{\rho}$ and $\mathbf{W}_{\rho}$ satisfy equation \eqref{egn; tangent map of orbit immersion, linear action} if and only if they coincide, and thus $T_{\rho}i$ is injective for all $\rho\in\mathcal{O}$.
\end{enumerate}
\qed

\end{proposition}

It is important to note that the topology  underlying the differential structure on $\mathcal{O}$ comes from the topology of $\gapp$ in the sense that a subset $U$ of the orbit is open iff $(\tau_{\rho}^{\Phi})^{-1}(U)$ is open in $\gapp$. 
In principle, this topology on $\mathcal{O}$ has nothing to do with the topology of $\mathcal{O}$ when thought of as a subset of $\stsp$ endowed with the relativised norm topology, or with the relativised weak* topology.
However, from proposition \ref{prop: canonical immersion of an orbit in the dual is smooth}, it follows that the map $l_{\mathbf{a}}\colon\,\mathcal{O}\lra\mathbb{R}$ is continuous for every $\mathbf{a}\in\appa_{sa}$.
Therefore, we may conclude that the topology underlying the homogeneous Banach manifold structure on $\mathcal{O} $ is stronger than the relativised weak* topology coming from $\appa^{*}_{sa}$.

\subsection{Density operators}\label{sec: density operators}

Similarly to what is done in subsection \ref{sebsec: positive trace-class operators}, we consider a  complex, separable Hilbert space $\hh$ and denote by $\appa$ the $W^{*}$-algebra $\bh$ of bounded, linear operators on $\hh$.
A normal state $\widetilde{\rho}$ of $\appa$ may be identified with a density operator on $\hh$, that is, a trace-class, positive semidefinite operator $\rho$ with unit trace, and the duality relation may be expressed by means of the trace operation
\be
\widetilde{\rho}(\mathbf{a})\,=\,\mathrm{Tr}\,\left(\rho\,\mathbf{a}\right)
\ee
for all $\mathbf{a}\in\appa=\mathcal{B}(\hh)$.

We will study the orbits of the group $\gapp$ of invertible, bounded linear operators in $\appa$ on the space $\stspn$ of normal states on $\appa$.
The analysis will be very similar to the one presented in subsection \ref{sebsec: positive trace-class operators}.

According to the spectral theory for compact operators (see \cite[ch. VII]{reed_simon-methods_of_modern_mathematical_physics_I_functional_analysis}), given a density operator $\rho$ on $\hh$, there is a decomposition $\mathcal{H}=\mathcal{H}_{\rho}\oplus\mathcal{H}_{\rho}^{\perp}$ and a countable orthonormal basis $\{|e_{j}\rangle,\,|f_{j}\rangle\}$ adapted to this decomposition such that $\rho$ can be written as
\be
\rho\,=\,\sum_{j=1}^{\mathrm{dim}(\mathcal{H}_{\rho})}\, p^{j}\,|e_{j}\rangle\langle e_{j}|\,,
\ee
with $p^{j}>0$ and $\sum_{j=1}^{\mathrm{dim}(\mathcal{H}_{\rho})}\, p^{j}=1$.
In general, we have four different situations:

\begin{enumerate}
\item $0<\mathrm{dim}(\mathcal{H}_{\rho})=N<\infty$;
\item $\mathrm{dim}(\mathcal{H}_{\rho})=\infty$ and $0<\mathrm{dim}(\mathcal{H}^{\perp}_{\rho})=M<\infty$;
\item $\mathrm{dim}(\mathcal{H}_{\rho})=\infty$ and $\mathrm{dim}(\mathcal{H}^{\perp}_{\rho})=0$
\item $\mathrm{dim}(\mathcal{H}_{\rho})=\mathrm{dim}(\mathcal{H}_{\rho}^{\perp})=\infty$,
\end{enumerate}
and we set
\be
\begin{split}
\stspn_{N}&\,:=\,\left\{\rho\in\stspn\,\,|\;0<\mathrm{dim}(\mathcal{H}_{\rho})=N<\infty\right\} \\ 
\stspn_{M}^{\perp}&\,:=\,\left\{\rho\in\stspn\,\,|\;\mathrm{dim}(\mathcal{H}_{\rho})=\infty\,\mbox{ and }\,0<\mathrm{dim}(\mathcal{H}^{\perp}_{\rho})=M<\infty\right\} \\ 
\stspn_{0}^{\perp}&\,:=\,\left\{\rho\in\stspn\,\,|\;\mathrm{dim}(\mathcal{H}_{\rho})=\infty\,\mbox{ and }\,\mathrm{dim}(\mathcal{H}^{\perp}_{\rho})=0\right\} \\ 
\stspn_{\infty}&\,:=\,\left\{\rho\in\stspn\,\,|\;\mathrm{dim}(\mathcal{H}_{\rho})=\mathrm{dim}(\mathcal{H}_{\rho}^{\perp})=\infty\right\} .
\end{split}
\ee
The subscript here denotes either the dimension of the space on which $\rho$  operates, or its codimension when the symbol $\perp$ is used.
Clearly, when $\mathrm{dim}(\hh)<\infty$, we have $\stspn_{N}=\emptyset$ for all $N>\mathrm{dim}(\hh)$, and $\stspn_{M}^{\perp}=\stspn_{0}^{\perp}=\stspn_{\infty}=\emptyset$.

\vsp

Propositions \ref{prop: isotropy subalgebra of Phi is the isotropy subalgebra of alpha plus the identity with real coefficients}, \ref{prop: isotropy group of Phi is Banach-Lie subgroup iff isotropy group of alpha is}, and \ref{prop: the isotropy subalgebra of a positive trace-class operator for the linear action is complemented} imply that the isotropy subgroup $\gapp_{\rho}$ of $\rho$ with respect to $\Phi$ is a Banach-Lie subgroup of $\gapp=\mathcal{GL}(\hh)$.
By adapting the proof of proposition \ref{prop: partial characterization of orbits of positive trace-class operators} in the obvious way, we may prove the following proposition:

\begin{proposition}\label{prop: partial characterization of orbits of density operators}
Let $\varrho_{0}$ and $\varrho_{1}$ be density operators on $\hh$, that is, positive, trace-class operators with unit trace.
Denote by $p_{0}^{j}$ and $p_{1}^{j}$ the $j$-th eigenvalue of $\varrho_{0}$ and $\varrho_{1}$, respectively.
If   $\hh_{\varrho_{0}}$ is isomorphic to $\hh_{\varrho_{1}}$, and $\hh_{\varrho_{0}}^{\perp}$ is isomorphic to $\hh_{\varrho_{1}}^{\perp}$, and if the following condition holds
\be 
\frac{p_{1}^{j}}{p_{0}^{j}}\,\leq\, C \,<\,\infty\;\;\;\;\;\forall j=1,...,N,
\ee
then,  the element $\gr$ given by 
\be\label{eqn: invertible element}
\gr\,:=\,\sum_{j=1}^{N}\,\sqrt{\frac{p^{j}_{1}}{p^{j}_{0}}}\,|e_{j}^{1}\rangle\langle e_{j}^{0}| + \sum_{k=1}^{M}\,|f_{k}^{1}\rangle\langle f_{k}^{0}|
\ee
is a bounded, invertible operator on $\hh$ such that
\be
\varrho_{1}\,=\,\frac{\gr\,\varrho_{0}\,\gr^{\dagger}}{Tr(\gr\,\varrho_{0}\,\gr^{\dagger})}\,.
\ee
\end{proposition}

Clearly, the assumptions in proposition \ref{prop: partial characterization of orbits of density operators} are always satisfied if $\varrho_{0}$ and $\varrho_{1}$ are finite-rank operators with the same rank.

\vsp

Now, we want to   write down  a tangent vector $\mathbf{V}_{\rho}$ at $\rho\in\mathcal{O}$, where $\mathcal{O}$ is any of the orbits of $\gapp$ inside the space of density operators.
At this purpose, we consider the canonical immersion $i\colon\mathcal{O}\lra\appa^{*}_{sa}$, and we recall equation \eqref{eqn: tangent vector at a point in the orbit of Phi}, from which it follows that 
\be
T_{\rho}i(\mathbf{V}_{\rho})(\mathbf{b})\,=\,\rho\left(\mathbf{a}^{\dagger}\,\mathbf{b}\,+\,\mathbf{b}\,\mathbf{a}\right)\,-\,\rho(\mathbf{b})\,\rho\left(\mathbf{a}^{\dagger}\,+\,\mathbf{a}\right)\;\;\;\forall\mathbf{b}\in\appa,
\ee
where $\mathbf{a}$ is an arbitrary element in $\appa=\bh$.
Clearly, different choices of $\mathbf{a}$ may lead to the same $T_{\rho}i(\mathbf{V}_{\rho})$.
Then, writing $\mathbf{a}=\mathbf{x} + \imath\mathbf{y}$ with $\mathbf{x},\mathbf{y}\in\appa_{sa}$, we have
\be
T_{\rho}i(\mathbf{V}_{\rho})(\mathbf{b})\,=\,\mathrm{Tr}\left(\left(\left\{\rho,\,\mathbf{x}\right\} -\imath\,\left[\rho,\, \mathbf{y}\right]\right)\,\mathbf{b}\right) - \mathrm{Tr}(\rho\,\mathbf{b})\,\mathrm{Tr}\left(\left\{\rho,\,\mathbf{x}\right\} \right)\;\;\;\forall\mathbf{b}\in\appa,
\ee
with $\{\cdot,\,\cdot\}$ and $[\cdot,\,\cdot]$ the anticommutator and the commutator in $\bh$, respectively.

\subsection{Faithful, tracial state}\label{subsec: faitfhul, tracial state}

Similarly to what is done in \ref{subsec: faithful, finite trace}, we consider $\appa$ to be a unital $C^*$-algebra with a faithful, tracial state $\tau$, that is, $\tau$ is a faithful, state on $\appa$ such that
\be
\tau(\mathbf{a}\,\mathbf{b})\,=\,\tau(\mathbf{b}\,\mathbf{a})\;\;\;\forall\,\,\mathbf{a},\mathbf{b}\in\appa .
\ee
In particular, if $\appa$ is Abelian, then every faithful, state is a faithful, tracial state.

The result of subsection \ref{subsec: faithful, finite trace} and proposition \ref{prop: isotropy group of Phi is Banach-Lie subgroup iff isotropy group of alpha is} allow us to conclude that the orbit $\mathcal{O}^{\tau}$  of $\gapp$ through $\tau$ by means of $\Phi$ (see equation \eqref{eqn: non convex action of the group of invertibles on the space of states}) is a homogeneous Banach manifold of $\gapp$.
Furthermore, it is immediate to check that there is a bijection between $\mathcal{O}^{\tau}$ and the set of positive, invertible elements in $\appa$ with unit trace, that is, elements of the form $\frac{\gr\,\gr^{\dagger}}{\tau(\gr\,\gr^{\dagger})}$ with $\gr\in\gapp$.
If $\appa$ is finite-dimensional, then $\mathcal{O}^{\tau}$ coincides with the whole space of faithful, tracial states, and, if $\appa$ is finite-dimensional and Abelian (i.e., $\appa\cong\mathbb{C}^{n}$ for some $n\in\mathbb{N}$), then $\mathcal{O}^{\tau}$ may be identified with the open interior of the $n$-dimensional simplex.
Note that points in the orbit through $\tau$ need not be tracial state when $\appa$ is non-Abelian (e.g., when $\appa=\mathcal{B}(\hh)$ with $\mathrm{dim}(\hh)<\infty$ and $\tau$ the maximally mixed state).

\vsp

Now, we want to explore the example given by the Abelian $W^*$-algebra $\appa=L^{\infty}(\mathcal{X},\nu)$, where $(\mathcal{X},\Sigma,\nu)$ is a probability space (see \cite[p. 109]{takesaki-theory_of_operator_algebra_I}), and the support of $\nu$ is the whole $\mathcal{X}$.
The sum, multiplication and involution in $\appa=L^{\infty}(\mathcal{X},\nu)$ are defined as for the $C^*$-algebra of complex-valued, bounded, continuous functions on a Hausdorff topological space, but the norm is  given by
\be
||f||\,=\,\inf\left\{C\geq 0\,|\;|f(x)|\leq C\:\mbox{ for } \nu-\mbox{almost every }  x\right\}\,.
\ee
In this case, the pre-dual space $\appa_{*}$  may be identified with $L^{1}(\mathcal{X},\nu)$ by means of the duality
\be
\langle f,\xi\rangle\,=\,\int_{\mathcal{X}}\,f(x)\,\xi(x)\,\mathrm{d}\nu(x),
\ee
while the dual space $\appa^{*}$ may be identified with the space $BV(\Sigma,\nu)$ of complex-valued, finitely-additive, bounded functions on $\Sigma$ which vanish on every locally $\nu$-null set (see \cite[p. 116]{takesaki-theory_of_operator_algebra_I} for the explicit construction of the Banach space structure on $BV(\mathcal{X},\nu)$) by means of the duality
\be
\mu(f)\,=\,\int_{\mathcal{X}}\,f(x)\,\mathrm{d}\mu(x).
\ee
The space $\stsp$ of states is then the space of normalized, positive, finitely-additive, bounded functions on $\Sigma$.
When $\mu$ is a normal state, there exists $\tilde{\mu}\in\appa_{*}\cong L^{1}(\mathcal{X},\nu)$ such that  
\be
\int_{\mathcal{X}}\,f(x)\,\mathrm{d}\mu(x)\,=\,\mu(f)\,=\,\langle f,\tilde{\mu}\rangle\,=\,\int_{\mathcal{X}}\,f(x)\,\tilde{\mu}(x)\,\mathrm{d}\nu(x)
\ee
for all $f\in\appa$.
Clearly, the function $\tilde{\mu}$ is $\mu$-integrable, non-negative, and such that 
\be
\int_{\mathcal{X}}\,\tilde{\mu}(x)\,\mathrm{d}\nu(x)\,=\,1.
\ee 
Consequently, every normal state $\mu$ determines a probability measure on $(\mathcal{X},\Sigma)$ which is absolutely continuous with respect to $\nu$, and has $\tilde{\mu}$ as its Radon-Nikodym derivative.

The action of $\gapp$ on the normal state $\mu$ is easily written as
\be
\left(\Phi(\gr,\,\mu)\right)(f)\,=\,\frac{\int_{\mathcal{X}}\,|\gr(x)|^{2}\,f(x)\,\mathrm{d}\mu(x)}{\int_{\mathcal{X}}\,|\gr(y)|^{2}\,\mathrm{d}\mu(y)}\,\equiv\, \int_{\mathcal{X}}\,\alpha_{\gr}^{\nu}(x)\,f(x)\,\mathrm{d}\mu(x)\,,
\ee
where $f$ is in $\appa=L^{\infty}(\mathcal{X},\nu)$, and $\alpha_{\gr}^{\nu}(x)$ is the strictly positive, $\mu$-integrable function
\be
\alpha_{\gr}^{\mu}(x)\,=\,\frac{|\gr(x)|^{2}}{\int_{\mathcal{X}}\,|\gr(y)|^{2}\,\mathrm{d}\mu(y)}\,
\ee
such that $\mu(\alpha_{\gr}^{\mu})=1$.
If $\mu$ is faithful, the orbit $\mathcal{O}^{\mu}$ through $\mu$ is a smooth, homogeneous Banach manifold for $\gapp$, and a point in $\mathcal{O}^{\mu}$ is a probability measure $\mu_{\gr}$ which is mutually absolutely continuous with respect to $\mu$ with $\alpha_{\gr}^{\mu}\in\appa=L^{\infty}(\mathcal{X},\nu)$ as its Radon-Nikodym derivative.

In particular, if we take $\nu$ as the reference faithful, normal state, we have that the orbit $\mathcal{O}$ containing $\nu$ is given by all the probability measures on $(\mathcal{X},\Sigma)$ that are mutually absolutely continuous with respect to $\nu$, and with a Radon-Nikodym derivative which is a  strictly positive function in $\appa=L^{\infty}(\mathcal{X},\nu)$ integrating to $1$ with respect to $\nu$.
Therefore, the set
\be
\mathscr{M}_{\nu}\,:=\,\left\{f\in\,L^{\infty}(\mathcal{X},\nu)\,|\;\;f(x)>0\,\,\nu-a.s. \,,\;\int_{\mathcal{X}}\,f(x)\,\mathrm{d}\nu(x)=1\,\right\}
\ee
is a homogeneous Banach manifold of the Banach-Lie group $\gapp$ of invertible elements in $L^{\infty}(\mathcal{X},\nu)$.

\section{Concluding remarks}\label{sec: concluding remarks}

In this work, we presented a preliminary analysis concerning two possible actions of the  Banach-Lie group $\gapp$ of invertible elements in a unital $C^{*}$-algebra $\appa$ on the continuous, self-adjoint linear functionals in $\appa^{*}_{sa}$.
Specifically, we analysed a linear action $\alpha$ of $\gapp$ on $\appa^{*}_{sa}$ which is smooth and preserves the positivity and the normality of the linear functionals on which it acts.
In the case where $\appa$ is the algebra $\bh$ of bounded linear operators on a complex, separable Hilbert space $\hh$, we were able to prove that all the orbits passing through normal, positive linear functionals (positive trace-class operators on $\hh$) are smooth, homogeneous Banach manifolds of $\gapp=\mathcal{GL}(\hh)$ with respect to the action $\alpha$.
Furthermore, we gave sufficient conditions for two normal, positive linear functionals to belong to the same orbit.
If $\appa$ admits a faithful, finite trace $\tau$, then we proved that the orbit through $\tau$ is a smooth, homogeneous Banach manifold of $\gapp$.

The action $\alpha$ does not preserve the space of states $\stsp$ on $\appa$.
Consequently, we provided a sort of deformation of $\alpha$, denoted by $\Phi$, which allows us to overcome this problem.
However, $\Phi$ turns out to be an action of $\gapp$ which is well-defined only on the space of states $\stsp$, and, in general, it does not preserve the convex structure of $\stsp$.
The subgroup of unitary elements in $\gapp$ is the maximal subgroup such that the restriction of $\Phi$ preserves convexity.
Since $\stsp$ lacks a differential structure as a whole, it is meaningless to speak of the smoothness of $\Phi$, nevertheless, an orbit $\mathcal{O}$ of $\Phi$ may still inherit the structure of smooth homogeneous Banach manifold if the isotropy subgroup of an element (and thus of every element) in $\mathcal{O}$ is a Banach-Lie subgroup of $\gapp$.
At this purpose, we analysed the case where $\appa=\bh$ mentioned before, and we proved that the orbits through normal states (density operators on $\hh$) are indeed smooth homogeneous Banach manifolds for $\gapp=\mathcal{GL}(\hh)$  with respect to the action $\Phi$.
Similarly to what we obtained for the action $\alpha$ in the case of normal positive functionals on $\appa=\bh$, we gave sufficient conditions for two normal states to belong to the same orbit.
Furthermore, if $\appa$ admits a faithful, tracial state $\tau$, then we proved that the orbit of $\gapp$ through $\tau$ by means of $\Phi$ is a smooth, homogeneous Banach manifold of $\gapp$.
In particular, if $\appa$ is finite-dimensional, the orbit through $\tau$ coincides with the space of faithful states on $\appa$, while, if $\appa$ is finite-dimensional and Abelian, the orbit through $\tau$ may be identified with the open interior of the finite-dimensional simplex.
Note that points in the orbit through $\tau$ need not be tracial state when $\appa$ is non-Abelian, e.g., when $\appa=\mathcal{B}(\hh)$ with $\mathrm{dim}(\hh)<\infty$ and $\tau$ the maximally mixed state.

In the finite-dimensional case when $\appa=\mathcal{B}(\hh)$ with $\mathrm{dim}(\hh)<\infty$, the space of faithful states (invertible density operators) may be identified with a smooth, open submanifold of the affine hyperplane of the self-adjoint operators with unit trace, and numerous constructions related with classical information geometry have been adapted to this quantum case.
However, a straightforward extension of this formalism to the infinite dimensional case is not possible because no density operator can be invertible in this case, and this prevents the possibility of endowing the set of faithful (normal) states with a smooth manifold structure as it is done in the finite-dimensional case.
As noted in the introduction, in the finite-dimensional case it is also known that the manifold structure on faithful normal states admits a compatible transitive action of the group of invertible elements making it a smooth homogeneous space.
Clearly, these two manifold structures are completely equivalent in the finite-dimensional case.
A relevant conclusion that can be drawn from the results of this manuscript is that, in the infinite-dimensional case, it is  possible to define a smooth homogeneous structure on the orbit of $\gapp=\mathcal{GL}(\hh)$ through any given faithful normal state.
However, it is still an open question if there is only one such orbit as in the finite-dimensional case, and we plan to address this delicate issue in a future work.

The content of this work should be thought of as a preliminary step toward the generalization of the methods of quantum information geometry to the infinite-dimensional case, and, as such, it is far from being complete.  
For instance, the problem of characterizing the isotropy subgroups for other types of $C^{*}$-algebras other than those considered here in order to understand if the associated orbit (with respect to $\alpha$ or $\Phi$) is a smooth homogeneous Banach manifold for $\gapp$ is still open.
Then,  it would be relevant to analyse the smoothness of well-known informational quantities (e.g., quantum relative entropies) with respect to the smooth structures introduced in this work, so that, if smoothness is assured, we may proceed to analyse the statistical structures they give rise to in the infinite-dimensional case.  
Relevant examples would be given by the von Neumann-Umegaki relative entropy, and by the Bures distance function (quantum fidelity).

Other geometrical structures that will emerge naturally for certain states, like a complex structure or a symplectic structure \cite{beltita_ratiu-symplectic_leaves_in_real_banach_lie-poisson_spaces}, \cite{beltita_ratiu-unitary}, induced from the Banach-Lie group of  invertible elements, will be studied in a future contribution in relation with the well-known informational metrics. 

For this purpose, it may turn out to be helpful to analyse the smooth structure on the homogeneous Banach manifolds in terms of the smooth subalgebra of $\appa$  determined by the isotropy subgroup of the state $\rho$  ``labelling'' the homogeneous Banach manifold \cite{Br86}.   
Clearly, the smooth structure of the algebra will be related to the smooth structure of the orbit and it will provide a new insigh in the structure of the $C^*$-algebra obtained from an action of a Lie group as indicated for instance in  \cite{Co01}.
We plan to address these and related issues in future publications.

\section*{Acknowledgements}

A.I. and G.M. acknowledge financial support from the Spanish Ministry of Economy and Competitiveness, through the Severo Ochoa Programme for Centres of Excellence in RD (SEV-2015/0554).
A.I. would like to thank partial support provided by the MINECO research project MTM2017-84098-P and QUITEMAD++, S2018/TCS-­4342.
G.M. would like to thank the support provided by the Santander/UC3M Excellence Chair Programme 2019/2020.

\appendix

\section{C*-algebras  and states}\label{app: C*-algebras preliminaries}

Let $(\appa,\,+,\,||\cdot||,\,\cdot)$ be a Banach algebra, that is, a Banach space endowed  with a (not necessarily commutative) multiplication operation which is continuous with respect to the norm topology and is such that 
\be
||\mathbf{a}_{1}\,\mathbf{a}_{2}||\,\leq\,||\mathbf{a}_{1}||\,\,||\mathbf{a}_{2}||
\ee
for every $\mathbf{a}_{1}$ and $\mathbf{a}_{2}$ in $\appa$.
Suppose that there is a linear anti-isomorphism $\dagger$ on $\appa$, called involution, such that $(\mathbf{a}^\dagger)^\dagger = \mathbf{a}$ and:
\be
(\mathbf{a}_{1}\,\mathbf{a}_{2})^{\dagger}\,=\,\mathbf{a}_{2}^{\dagger}\,\mathbf{a}_{1}^{\dagger}
\ee
for every $\mathbf{a}_{1}$ and $\mathbf{a}_{2}$ in $\appa$.
If the pentuple $(\appa,\,+,\,||\cdot||,\,\cdot,\,\dagger)$ satisfies the compatibility condition between the norm, the multiplication and the involution given by 
\be
||\mathbf{a}||^{2}\,=\,||\mathbf{a}^{\dagger}\,\mathbf{a}||
\ee
for every $\mathbf{a}$ in $\appa$, then $(\appa,\,+,\,||\cdot||,\,\cdot,\,\dagger)$ is called \grit{$C^{*}$-algebra}. 
In the following, we will avoid the notation $(\appa,\,+,\,||\cdot||,\,\cdot,\,\dagger)$ to denote a $C^{*}$-algebra, and will simply write $\appa$ because, hopefully, all the operations will be clear from the context.

An element $\mathbf{a}\in\appa$ is called \grit{self-adjoint} if $\mathbf{a}=\mathbf{a}^{\dagger}$, and we denote by $\appa_{sa}$ the space of self-adjoint elements in $\appa$.
An element $\mathbf{a}\in\appa$ is called \grit{positive} if it can be written as 
\be
\mathbf{a}=\mathbf{b\,b}^{\dagger}
\ee
for some $\mathbf{b}\in\appa$.
Without loss of generality, we may take $\mathbf{b}$ to be self-adjoint (see \cite[p. 33]{bratteli_robinson-operator_algebras_and_quantum_statistical_mechanics_1}).
The space of positive elements in $\appa$ is denoted by $\appa_{+}$ and it is easy to see that it is a cone in $\appa$.

Let $\appa$ be a possibly infinite-dimensional, unital $C^{*}$-algebra, that is, a $C^{*}$-algebra with a multiplicative identity element denoted by $\mathbb{I}$.
Let $\appa^{*}$ be the topological dual of $\appa$, that is, the space of complex-valued, continuous linear functions on $\appa$.
We denote by $\mathcal{T}$ the Banach space topology (norm topology) on $\appa^{*}$ induced by the canonical norm
\be
\left|\left|\omega\right|\right|:=\sup_{||\mathbf{a}||=1}\, |\omega(\mathbf{a})|\,.
\ee
A linear functional $\omega\in\appa^{*}$ is called \grit{self-adjoint} if
\be
\omega(\mathbf{a}^{\dagger})\,=\,\overline{\omega(\mathbf{a})}
\ee
so that $\omega$ takes real values when evaluated on self-adjoint elements in $\appa$.
We denote by  $\appa^{*}_{+}$ be the cone of positive linear functionals on $\appa$, that is, the set of all $\omega\in\appa^{*}$ such that $\omega(\mathbf{a}^{\dagger}\,\mathbf{a})\geq 0$ for every $\mathbf{a}\in\appa$.
According to  \cite[p. 49]{bratteli_robinson-operator_algebras_and_quantum_statistical_mechanics_1}, positive linear functionals are self-adjoint, and, given $\omega\in\appa^{*}_{+}$, it holds  $||\omega||=\omega(\mathbb{I})$ where $\mathbb{I}$ is the identity  of $\appa$.
We denote by $\stsp$ the space of states of $\appa$, that is, the set of all $\omega\in\appa^{*}_{+}$ such that $||\omega ||\,=\, 1$. 
A state $\omega\in\stsp$ is called  faithful if $\omega(\mathbf{a})> 0$ for all $\mathbf{a} \in \appa_+$.
A state $\omega\in\stsp$ is called pure if, given $\xi\in \appa^{*}_{+}$, then $(\omega - \xi)$ is in $\appa^{*}_{+}$ only if $\xi=\lambda\omega$ with $0\leq\lambda\leq 1$.
The set of pure states is denoted by $\mathscr{P}$, and, according to \cite[p. 53]{bratteli_robinson-operator_algebras_and_quantum_statistical_mechanics_1}, the space of states $\stsp$  is  a convex, weak* compact convex subset of $\appa^{*}$, and $\mathscr{P}$ is the set of extremal points of $\stsp$, that is, $\stsp$ is the  weak* closure of the convex envelope of $\mathscr{P}$.
Note that the weak* compactness of $\stsp$ depends on the fact that $\appa$ has an identity element.

Given a positive linear functional $\omega\in\appa^{*}_{+}$, it is always possible to build a triple $(\hh_{\omega},\,\pi_{\omega},\,|\psi_{\omega}\rangle)$ where $\hh_{\omega}$ is a possibly infinite-dimensional complex Hilbert space, $\pi_{\omega}$ is a *-representation of $\appa$ in $\mathcal{B}(\hh_{\omega})$, and $|\psi_{\omega}\rangle$  is a nonzero vector in $\hh_{\omega}$ such that
\be
\omega(\mathbf{a})\,=\,\langle\psi_{\omega}|\pi_{\omega}(\mathbf{a})|\psi_{\omega}\rangle
\ee  
for all $\mathbf{a}\in\appa$, and such that 
\be
H_{\omega}\,:=\,\left\{|\psi\rangle\,\in\,\hh_{\psi}\;\colon\;\;\;\exists\,\,\mathbf{a}\in\appa\;\mbox{ such that }\;|\psi\rangle\,=\,\pi_{\omega}(\mathbf{a})|\psi_{\omega}\rangle\right\}\,
\ee
is a dense subset in $\hh_{\omega}$.
This construction is referred to as the GNS construction associated with $\omega$ (see \cite[sec. II.6.4]{blackadar-operator_algebras_theory_of_c*-algebras_and_von_neumann_algebras},  \cite[ch. 2.3]{bratteli_robinson-operator_algebras_and_quantum_statistical_mechanics_1} and \cite[ch. 4.5]{kadison_ringrose-fundamentals_of_the_theory_of_operator_algebras_I} for more details).
Note that the GNS construction associated with $\omega$ is unique up to unitary isomorphisms, and there is a one-to-one correspondence between positive linear functionals in $\appa^{*}_{+}$ and the $*$-representations of $\appa$ with a specified cyclic vector.
Furthermore, the GNS representation $\pi_{\omega}$ is irreducible if and only if $\omega$ is a pure state (see \cite[p. 57]{bratteli_robinson-operator_algebras_and_quantum_statistical_mechanics_1}) and two irreducible GNS  representations $\pi_{\omega}$ and $\pi_{\rho}$ are unitarily equivalent if $||\omega-\rho||<2$ (see \cite[p. 551]{glimm_kadison-unitary_operators_in_cstar_algebras}).

A $W^{*}$-algebra $\appa$ is a $C^{*}$-algebra which is the Banach dual of a Banach space $\appa_{*}$ called the  predual of  $\appa$.
According to \cite[p. 30]{sakai-C_star_and_W_star_algebras}, the predual $\appa_{*}$ is unique up to isometric isomorphisms.
We denote by $\langle\cdot,\cdot\rangle$ the canonical pairing between $\appa_{*}$ and $\appa$, that is, the action of $\mathbf{a}\in\appa$ on $\xi\in\appa_{*}$ when the former is thought of as a continuous linear functional on $\appa_{*}$, reads 
\be
\langle \xi,\mathbf{a} \rangle = \mathbf{a}(\xi).
\ee
There is a natural immersion $\mathrm{i}$ of $\appa_{*}$ into its double (topological) dual $\appa^{*}$ given by $\xi\,\mapsto\;\mathrm{i}(\xi)$ with $\left(\mathrm{i}(\xi)\right)(\mathbf{a})\,=\,\langle\mathbf{a},\,\xi\rangle$, 
and, by the very definition of the weak* topology on $\appa$, it is clear that a linear functional $\widetilde{\xi}$ in $\appa^{*}$ is continuous with respect to the weak* topology on $\appa$ if and only if there is an element $\xi\in\appa_{*}$ such that $\mathrm{i}(\xi)=\widetilde{\xi}$.
Linear functionals of this type are called \grit{normal}, and the set of normal states on $\appa$ is denoted by $\stspn$.
In the finite dimensional case, every  $C^{*}$-algebra is a $W^{*}$-algebra, every linear functional $\xi\in\appa^{*}$ is normal, and the immersion $\mathrm{i}$ is actually an isomorphism.
This is no longer true in the infinite dimensional case.

\section{Banach-Lie groups and homogeneous spaces}\label{app: banach-lie groups and homogeneous spaces}

In this section we will assume familiarity with the notions of \grit{real}, smooth Banach manifolds, smooth maps between (smooth) Banach manifolds, and Banach-Lie groups.
As stated at the end of the introduction, the main references concerning the infinite-dimensional differential geometry of Banach manifolds and Banach-Lie groups are \cite{abraham_marsden_ratiu-manifolds_tensor_analysis_and_applications,bourbaki-groupes_et_algebres_de_lie,chu-jordan_structures_in_geometry_and_analysis,lang-fundamentals_of_differential_geometry,upmeier-symmetric_banach_manifolds_and_jordan_calgebras},  however, we think it is useful to recall here some notions regarding  Banach-Lie subgroups of Banach-Lie groups.

According to \cite[p. 96 and p. 114]{upmeier-symmetric_banach_manifolds_and_jordan_calgebras}, every closed subgroup $K$ of a given Banach-Lie group $G$ is a Banach-Lie group with respect to a unique Hausdorff topology in $K$ such that the closed real subalgebra
\be\label{eqn: banach-lie algebra of banach-lie subgroup}
\mathfrak{k}=\left\{\mathbf{a}\in\mathfrak{g}\,\colon \exp(t\mathbf{a})\in K\;\forall t\in\mathbb{R}\right\}\,
\ee
of the Lie algebra $\mathfrak{g}$ of $G$ is the Lie algebra of $K$.
In general, the Hausdorff topology on $K$ does not coincide with the relative topology inherited from the norm topology of $G$.
A subgroup $K$ of a Banach-Lie group $G$ which is also a Banach submanifold of $G$ is called a \grit{Banach-Lie subgroup} of $G$.
In particular, a Banach-Lie subgroup $K$ of $G$ is closed, it is a Banach-Lie group with respect to the relative topology inherited from the topology of $G$, and its Lie algebra $\mathfrak{k}$ is given by equation \eqref{eqn: banach-lie algebra of banach-lie subgroup} (see \cite[p. 128]{upmeier-symmetric_banach_manifolds_and_jordan_calgebras}).

Note that, in the finite-dimensional case $\mathrm{dim}(G)<\infty$, it is always true that every closed subgroup of $G$ is  a Banach-Lie subgroup $G$.

An explicit characterization of Banach-Lie subgroups is given by the following proposition (see \cite[p. 129]{upmeier-symmetric_banach_manifolds_and_jordan_calgebras}).

\begin{proposition}\label{prop: characterization of Banach-Lie subgroups}
A closed subgroup $K$ of a Banach-Lie group $G$ is a Banach-Lie subgroup if and only if the closed subalgebra $\mathfrak{k}$  given by equation \eqref{eqn: banach-lie algebra of banach-lie subgroup} is a split subspace of the Lie algebra $\mathfrak{g}$ of $G$, and for every neighbourhood $U$ of $\mathbf{0}\in\mathfrak{k}$, we have that $\exp(V)$ is a neighbourhood of the identity element in $K$.
\end{proposition}

The importance of Banach-Lie subgroups  comes from the following theorem (see   \cite[p. 105]{bourbaki-groupes_et_algebres_de_lie} and \cite[p. 136]{upmeier-symmetric_banach_manifolds_and_jordan_calgebras}).

\begin{theorem}\label{thm: banach-lie subgroups and homogeneus spaces}
Let $K$ be a Banach-Lie subgroup of the Banach-Lie group $G$ with Lie algebra $\mathfrak{g}$. 
Then, the quotient space $M\equiv G/K$ carries the structure of an analytic Banach manifold such that the canonical projectio $\pi\colon G\lra M$ is an analytic submersion.
Writing $[h] = hK\equiv m\in M$ with $h\in G$, we have that the Lie group $G$ acts analytically on $M$ by the left translation given $L(\gr,m):=[\gr h] = g h K$.
The kernel of the tangent map $T\pi_{e}\colon \mathfrak{g}\lra T_{\pi(e)}M$ at the identity element $e\in G$  coincides with the closed, split subalgebra $\mathfrak{k}$ of equation \eqref{eqn: banach-lie algebra of banach-lie subgroup} so that $T_{\pi(e)}M\cong \mathfrak{g}/\mathfrak{k}$.
\end{theorem}

A smooth Banach manifold $M$ which is diffeomorphic with $G/K$ for some Banach-Lie group $G$ with Banach-Lie subgroup $K$ is called a \grit{homogeneous Banach manifold} of $G$.
In particular, let $G$ be a Banach-Lie group acting on a set $S$ by means of the left action $\alpha\colon G\times S\lra S$, let $\mathcal{O}$ be an orbit of $G$ in $S$ by means of $\alpha$, and let $s\in\mathcal{O}\subseteq S$.
The isotropy subgroup of $G$ at $s$ is the set
\be
G_{s}\,:=\,\left\{\gr\in G\,|\;\alpha(\gr,s)=s\right\}\,.
\ee
Then, the map $i_{s}\colon G/G_{s}\rightarrow \mathcal{O}$ given by
\be
[\gr]\mapsto i_{s}([\gr])=\alpha(\gr,s)
\ee
defines a bijection from $G/G_{s}$ to $\mathcal{O}$ for every $s\in\mathcal{O}$.
Consequently, if $G_{s}$ is a Banach-Lie subgroup of $G$, we may endow the orbit $\mathcal{O}$ containing $s$ with the structure of Banach manifold according to theorem \ref{thm: banach-lie subgroups and homogeneus spaces} so that $\mathcal{O}$ becomes an homogeneous space of $G$ for which the projection map $\tau_{s}\colon G\lra \mathcal{O}$, obtained composing the projection of $G$ onto $G/G_{s}$ with the map $i_{s}$ introduced above, is a surjective submersion.
Note that $S$ is just a set and no structural properties on the action $\alpha$ are  required.
In particular, if $G$ acts transitively on $S$, then $S$ itself is an homogeneous Banach manifold of $G$.

Given a Banach manifold $N$, a useful tool for determining if a map $\psi\colon \mathcal{O}\rightarrow N$ is smooth is given by the following proposition provided we make the identifications $M=\mathcal{O}$,  $K=G$ and $\phi=\tau_{s}$ for some $s\in\mathcal{O}$ (see \cite[p. 125]{upmeier-symmetric_banach_manifolds_and_jordan_calgebras}).

\begin{proposition}\label{prop: smoothness of maps from homogeneous space is related with  smoothness of maps from group}
Let $M,N,K$ be Banach manifolds, and consider the following commutative diagram: 

\begin{center}
\begin{tikzpicture} 
\draw [->] (0.5,0) -- (2.5,0);
\draw [->] (1.2,-2.1)  -- (0,-0.4);
\draw [->]  (1.8,-2.1)  -- (3,-0.4) ;
\node at  (0,0)   {$M$};
\node at  (3,0)   {$N$};
\node at  (1.5,-2.5)   {$K$};

\node [left] at (0.4,-1.2)   {$\phi$};
\node [right] at (2.6,-1.2)   {$\varphi$};
\node [above] at (1.5,0)   {$\psi$};
\end{tikzpicture}
\end{center}

Then, if $\phi$ is a surjective submersion and $\varphi$ is smooth, then $\psi$ is smooth. 
If $\varphi$ is also a submersion, then $\psi$ is also a submersion. 
\end{proposition}

\end{document}